\documentclass[useAMS, usenatbib]{mn2e}
\usepackage{lscape}
\usepackage{rotating}
\usepackage{amssymb}
\usepackage{gensymb}

\def\msun{\hbox{M$_\odot$}}

\def\t4{\hbox{t$_{\rm 4}$}}

\def\msunyr{\hbox{M$_\odot$yr$^{-1}$}}

\def\cm3{\hbox{cm$^{-3}$}}

\input psfig.sty
\input epsf.sty
\usepackage{graphicx}
\usepackage{epsfig}
\newcommand{\dist}{distribution}

\title[The cluster population of NGC 1566]
{Properties of the cluster population of NGC 1566 and their implications}
\author[Hollyhead et al.]{K. Hollyhead$^1$, A. Adamo$^2$, N. Bastian$^1$, M. Gieles$^3$, J. E. Ryon$^4$ \\
$^{1}$ Astrophysics Research Institute, Liverpool John Moores University, 146 Brownlow Hill, Liverpool L3 5RF, UK\\
$^{2}$ Department of Astronomy, Oscar Klein Centre, Stockholm University, AlbaNova, Stockholm SE-106 91, Sweden\\
$^{3}$ Department of Physics, University of Surrey, Guildford, GU2 7XH, UK\\
$^{4}$ Department of Astronomy, University of Wisconsin-Madison, 475 North Charter Street, Madison, WI 53706, USA\\
}
\date{Accepted. Received; in original form}
\pagerange{\pageref{firstpage}--\pageref{lastpage}}
\pubyear{2016}
\begin{document}
\maketitle
\label{firstpage}
\begin{abstract}
We present results of a photometric study into the cluster population of NGC 1566, a nearby grand design spiral galaxy, sampled out to a Galactocentric radius of $\approx 5.5$ kpc. The shape of the mass-limited age distribution shows negligible variation with radial distance from the centre of the galaxy, and demonstrates three separate sections, with a steep beginning, flat middle and steep end. The luminosity function can be approximated by a power law at lower luminosities with evidence of a truncation at higher luminosity. The power law section of the luminosity function of the galaxy is best fitted by an index $\approx -2$, in agreement with other studies, and is found to agree with a model luminosity function, which uses an underlying Schechter mass function. The recovered power law slope of the mass distribution shows a slight steepening as a function of galactocentric distance, but this is within error estimates. It also displays a possible truncation at the high mass end. Additionally, the cluster formation efficiency ($\Gamma$) and the specific U-band luminosity of clusters ($T_L(U)$) are calculated for NGC 1566 and are consistent with values for similar galaxies. A difference in NGC 1566, however, is that the fairly high star formation rate is in contrast with a low $\Sigma_{SFR}$ and $\Gamma$, indicating that $\Gamma$ can only be said to depend strongly on $\Sigma_{SFR}$, not the star formation rate.
\end{abstract}
\begin{keywords} 
galaxies: individual: ngc1566, galaxies: star clusters: general
\end{keywords}

\section{Cluster populations}
\label{sec:pops}

The study of stellar clusters is vital to the understanding of star formation and galaxy evolution, as current theory proposes that the majority of all stars form in groups or clustered environments (e.g. \citealt{hopkins13, kruijssen12}). A grouping is defined as any sized collection of stars, independent of whether or not the collection is gravitationally bound.  More specifically, a cluster refers to a bound grouping while associations are unbound groupings \citep{blaauw64, gieles11}. It is difficult to define a cluster at young ($<10$ Myr) ages, as without detailed kinematical information of all group members (and potentially the gas in the region as well) it is not possible to determine if the grouping is bound.  At young ages there exists a continuous distribution of structures, whereas at ages greater than 10~Myr, a bimodal distribution arises with bound (compact) and unbound (expanding associations) groups \citep{portzwart10}. 

Hence, an object can only definitely be defined and classified as a cluster once it is dynamically evolved. This is in agreement with hierarchical star formation models, which indicate that clusters are not well defined and unique objects at young ages \citep{bastian11}.

The problem with defining and determining a cluster from an unbound grouping also lies in the limited spatial resolution of the telescope and lack of information on stellar dynamics (e.g. \citealt{ba12}). Without knowledge of the stellar dynamics within clusters/groups, it is impossible to unambiguously determine whether the collection is bound or not. One simplistic way is to use the size of the grouping. As shown in the Small Magellanic Cloud, groupings with an effective radius above 6 pc rapidly decline in number as a function of age (t$^{-1}$), i.e., 90\% of groups get disrupted every decade in age, but for groups with sizes below 6 pc, there is a flat distribution suggesting little disruption \citep{portzwart10}. By limiting cluster population studies to systems containing dense, centrally concentrated groups with ages $>$ 10 Myr, issues with identifying bound structures can be resolved.

Much work has been done on cluster populations in several nearby galaxies, most notably on M83 and the Small and Large Magellanic Clouds. By studying the distributions of different cluster properties across the galaxy you can infer how the clusters formed and determine how  they evolve over time. 

\subsection{Cluster disruption}
\label{sec:disruption}

Clusters that survive any initial mass loss during formation will not survive indefinitely, as they will continue to dissolve through a combination of internal and external processes. This can be seen in the decrease in the number of clusters with increasing age  (e.g. \citealt{elmegreen10}). 

Gas loss at very young ages is expected to be entirely environmentally independent as this should be an internal process, and occurs on very short timescales (less than a crossing time). Early cluster evolution due to gas expulsion should additionally be mass independent, as violent relaxation is responsible for dynamical restructuring after rapid gas loss (e.g. \citealt{bastian06}).  One the other hand, the gas rich birth environment of clusters has been suggested to actually cause many of the young clusters to disrupt before they can migrate to areas that are more gas poor (e.g., \citealt{elmegreen10,kruijssen11}) in which case the local environment would play a strong role in the determination of the survival fraction of young clusters. Theoretically, disruption should depend on the cluster's initial mass and its environment, with clusters in weaker tidal fields or higher masses surviving longer  (e.g. \citealt{baumgardt03, gieles06c}).  However, empirical studies have resulted in two separate theories for the disruption process. 

The process of disruption over the lifetime of the clusters is more questionable. There are two main empirical scenarios to explain the process: MID (Mass Independent Disruption) and MDD (Mass Dependent Disruption). The age and mass distributions of clusters, as studied in this project, strongly depend on how disruption is modelled during the cluster lifetime. MDD has been evidenced and displayed via empirical studies \citep{bout03, lamers05} and N-body simulations \citep{baumgardt01, gieles04} where disruption is found to depend on the initial mass of the cluster and the environment within the galaxy. The timescale for disruption has been found to depend on the cluster mass as M$^\gamma$, where $\gamma$ varies slightly between different galaxies, with a mean value of $\gamma = 0.62$ \citep{bout03}. Simply, this indicates that higher mass clusters live longer. In this scenario, disruption is also dependent on environment due to tidal effects and the ambient density within the galaxy \citep{lamers05, lamers06}. 

The opposing idea to MDD has been proposed, claiming that disruption, up to an age of $\sim1$ Gyr, is independent of mass and galaxy environment \citep{whitmore07}. MID is based on studies of the Antennae \citep{fall05}, Large Magellanic Cloud \citep{chandar10a} and the central regions of M83 \citep{chand10}. According to these studies, independence of disruption from mass and environment results in a quasi-universal age and mass distribution with the number of clusters declining as a function of age as t$^{-1}$ \citep{whitmore07}. 

Potential models combining these two ideas have been explored and can be found to fit observed data for fast or slow disruption and disruption from internal mechanisms or outside influence (such as nearby clouds) with certain assumptions within reasonable limits \citep{elmegreen10}. Additionally, if tidal shocks are sufficiently strong, disruption may be mass independent (with an age distribution as $t^{-1}$) \citep{kruijssen11},  however there will still be a strong environmental influence. This indicates that fully constraining disruption mechanisms is a difficult process.

Age distributions have been widely studied for different galaxies and can provide insight into the process of disruption in a galaxy. The general shape of the distribution is a power law section with a steepening at high ages. The shape of the distribution is determined by the star formation history of the galaxy and the amount of disruption present. If approximated as a single power-law, log(dN/dt) $\propto$ t$^\zeta$, where studies to date have found $\zeta$ to vary between 0 and -1 (e.g. \citealt{fall05}, \citealt{gieles07}, \citealt{chand10}, \citealt{silva11}, \citealt{ba12}, \citealt{ryon14}).

Studies of M83 show environmental dependence in the age distribution, ranging from nearly flat ($dN/dt \sim t^0$) to relatively steep ($dN/dt \sim t^{-0.7}$) \citep{silva14} at different galactocentric distances. This was also found by \cite{chand14} (radially dependent index of the age distribution) who needed to invoke radially dependent differences in the cluster formation history in order to bring the age distributions in the inner and outer regions of the galaxy into agreement. 

\cite{ba12} found that both the MID and MDD scenarios could provide good fits to the observed age distributions of clusters in M83, however, disruption needed to be strongly dependent on environment.  Additionally, these authors found that the mass function for the clusters was truncated, and the truncation value depended on environment.  The truncation in the mass function was necessary to include in disruption analysis in order to explain the age distribution in the MDD framework.

\begin{figure}
\centering
\includegraphics[width=8.5cm]{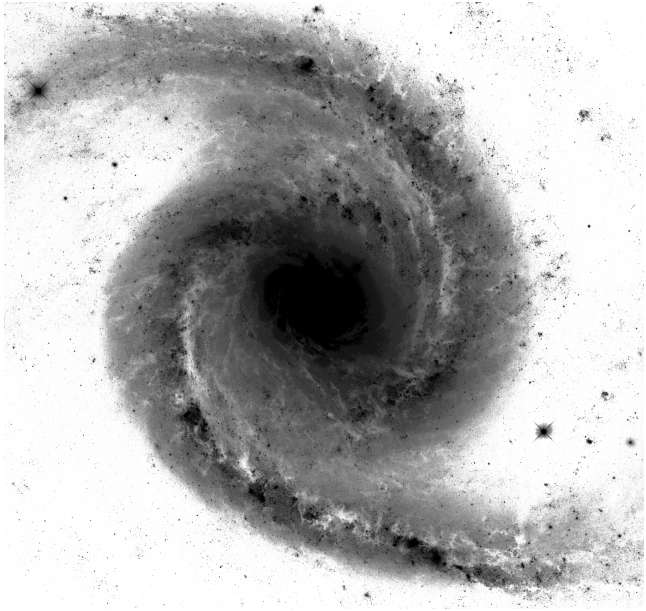}
\caption{Image of NGC 1566. The image was produced by equalizing and combining fits images of the galaxy in the B, V and I bands. The resulting image was then edited in the GNU GIMP image processing utility to change the image to black and white and invert the colours.}
\label{fig:galaxy}
\end{figure}

\subsection{Cluster population properties}
\label{sec:props}

The luminosity function (LF) of a cluster population provides insight into the mass function of the clusters. It is found to behave as $dN/dL \propto L^{-\alpha}$, although some deviations from a pure single valued power-law have been reported \citep{gieles06a}, including a steepening at the bright end. An example is the bend observed in the LF in the Antennae galaxies by \cite{whitmore99}, which can be fit with a double power law. When the power law section of the distribution is fit, $\alpha$ is usually found to be $\approx 2$, with small variations (e.g. \citealt{degrijs03, larsen02}). The luminosity function is related closely to the underlying mass function of the clusters, and the cluster initial mass function (CIMF). A direct comparison may not be made, however, as clusters of the same mass but varying ages will have different luminosities due to fading over time. A down-turn in the LF can result from a truncation at the high mass end of the CIMF (e.g. \citealt{gieles10}).  However, this truncation in the CIMF may be difficult to discern due to low numbers of massive clusters \citep{gieles06b}. 

The shape of the mass distribution has been questioned in a variety of studies, with some concluding that it is best fit with a pure power law (e.g. \citealt{bik03, whitmore10, chandar10a, chandar11}), while others say there must be a truncation in the high mass end, best fit by a Schechter function (e.g. \citealt{larsen09, gieles06b, maschberger09}). In the latter case, the fit is characterised by a power law at low masses with a truncation at the high mass end and a characteristic truncation mass that varies depending on galactic environment. The power law section of the mass distribution, where $dN/dM \propto M^{-\beta}$, has been found to be best fit in most cases with $\beta \approx 2$ (e.g. \citealt{zhang99, ba12}).

More recently, other galactic properties have been used to study cluster populations. One such property is the magnitude of the brightest cluster in the V-band of a population, $M_V^{brightest}$ \citep{larsen02, whitmore03}. There is an observed relation between this quantity and the star formation rate (SFR) of the galaxy which is interpreted to be due (mainly) to a size-of-sample effect (see \cite{adamo16} for a recent review).

$\Gamma$, or the cluster formation efficiency (CFE),  is the fraction of stars that form within clusters in a given environment/galaxy (see \cite{bastian08}).  \cite{kruijssen12} presented a model that relates the CFE to the formation process of clusters, so when local density distributions are taken into account, the CFE should scale with the gas surface density of the galaxy, leading to a decrease with distance from the centre of any individual (spiral) galaxy. The cluster formation efficiency should also then scale with the surface density of star formation ($\Sigma_{SFR}$) in the galaxy via the Schmidt-Kennicutt law \citep{kennicutt98, kennicutt12}, which has been observed in many galaxies to date (e.g. \citealt{goddard10, ryon14, adamo15}). This indicates that $\Gamma$ can be used to probe the effect of galactic environment on cluster population properties. Additionally, a variation in CFE with distance from the galactic centre has been found for M83 \citep{silvavilla13, adamo15}. 

Another correlation between cluster population properties and their host galaxy has been investigated by \cite{larsen00}, where they showed that the percentage of total U band luminosity of a galaxy contained within young massive clusters (YMCs) ($T_L(U)$) shows a relationship with several host qualities including $\Sigma_{SFR}$ and the density of HI emission. $T_L(U)$ can also provide insight into the environmental dependence of cluster populations in the galaxy \citep{larsen02}. As U band luminosity traces young stellar populations, $T_L(U)$ can be related to star formation and consequently, CFE. The relationship between $T_L(U)$ and $\Sigma_{SFR}$ of the galaxy indicates that the amount of star formation occurring in clusters increases with $\Sigma_{SFR}$, and so CFE should also increase, though the exact relation between the CFE and $T_L(U)$ has not yet been quantified. $T_L(U)$ is found to range from $\approx$ 0.1 - 15 in galaxies with a small cluster population to merging systems, respectively \citep{larsen00}. 

\subsection{Testing the theories}
\label{sec:test}

To address the open questions discussed in the previous section, we study the properties of the cluster population of NGC 1566, the brightest member in the Dorado group of galaxies, as shown in Fig.~\ref{fig:galaxy}. It is a face-on spiral Seyfert galaxy \citep{vauc73} at a distance of $\sim17$ Mpc \citep{karachen96}, which makes it ideal for identifying and studying its cluster population. At this distance individual clusters can still be resolved using HST data meaning photometry can yield integrated magnitudes for each cluster. Extensive HST data is already available for NGC 1566 on the Hubble Legacy Archive, covering the inner $\sim5$~kpc in many bands from the UV to the optical. This galaxy has also been studied in HI, H$\alpha$, CO, X-ray and radio continuum \citep{kilborn05, korchagin00}.

We have constructed a catalogue of clusters within the galaxy and obtained photometry in a variety of photometric bands using HST data. We fit simple stellar population models to the photometry to obtain ages, masses and extinctions for each of the clusters and then inspect the age and mass distributions of clusters in different areas of the disc. Throughout this work we use three separate radial bins to investigate changes in properties with environment. These bins are approximately equal in cluster number and are arranged from 0-3.3 kpc, 3.3-4.7 kpc and 4.7 kpc to the edge of the image for radial bins 1, 2 and 3 respectively. A variable age distribution could also indicate an environmental dependence in the disruption mechanism. We also compare our fits for the mass and age distributions for studies of other galaxies.

In \S~\ref{sec:obs} we discuss the observations and techniques we used to analyse the HST images of the galaxy. \S~\ref{sec:ccplot}, \S~\ref{sec:lumfunc}, \S~\ref{sec:agemass} discuss our results for the observations from NGC 1566, while \S~\ref{sec:models} and \S~\ref{sec:gamma} discuss comparisons of our data with models. \S~\ref{sec:discussion} provides an overview of the results and our conclusions.

\section{Observations and Techniques}
\label{sec:obs}

\subsection{Data and photometry}
\label{sec:data}

The images used in this study were taken from the Hubble Legacy Archive (HLA), having previously been fully reduced and drizzled with exposures covering the inner regions of NGC 1566 (out to $\approx 5.5$ kpc) over a range of wavelengths, assuming a distance modulus of $m-M \approx 31.2$. The galaxy is part of the Legacy ExtraGalactic UV Survey (LEGUS, \cite{calzetti15}, HST project number GO-13364) and as such has complete and homogeneous imaging coverage in the UV (F275W), U (F336W), B (F438W), V (F555W) and I (F814W) bands obtained using WFC3. No conversion was made to the Cousins-Johnson filter system, but the central wavelengths of these bands are approximately equal to the WFC3 bands, so we use this nomenclature for simplicity. 

No prior catalogue was available for this galaxy, so we carried out our own photometry on the images. We used Source Extractor \citep{sextractor} within the Gaia package \citep{gaia} included in the Starlink software to locate potential clusters throughout the galaxy using the V band image (the band where most clusters should be visible). No limiting parameters were applied to the source extraction procedure to minimise the number of clusters unintentionally omitted from the detection. Additionally, the catalogue would undergo extensive refinement at a later stage, so any unreliable sources or false detections would later be removed. $\sim20000$ sources were identified. The resulting coordinates were then used to carry out aperture photometry in the {\scshape daophot} package in {\scshape iraf}. Magnitudes for each source were obtained using apertures of 1, 3 and 5 pixels (corresponding to 0.04, 0.12 and 0.2 arcseconds respectively, or 3.3, 9.9 and 16.5 pc), with sky background annuli between 15 and 17 pixels (0.60 and 0.68 arcseconds or 49.5 and 56.0 pc respectively) for the UV, U, B, V, and I bands.

\subsection{Catalogue refinement}
\label{sec:catalogue}

The source catalogue included many false detections and unwanted objects such as stars or possibly background galaxies. The first cut made to the objects to reduce the number of false detections involved removing any objects that were not sufficiently detected in the U,  B, V and I bands. Any objects without photometric magnitudes in all of these bands were removed.  

Similar to the work done by \cite{chand10} and again by \cite{ba12} and \cite{silva14} for M83, we used concentration indices (CIs) to refine the sample further. A CI is a measurement of how centrally concentrated emission is for a source. This value helps distinguish between field stars and clusters; stars should be highly centrally concentrated as point sources, whereas clusters have extended emission. 

Aperture corrections were calculated in each of the bands to account for flux missed by using a small aperture. This process involved visually selecting 30 sources that passed our selection criteria and were isolated in a variety of locations in the galaxy. Photometry was carried out in {\sc iraf} for each of the sources with apertures from 1-15 pixels at 1 pixel intervals. These magnitudes were then used to create a radial profile for each cluster in each wavelength. Ideally the profiles rise rapidly from the centre of the cluster for the first few pixels then begin to flatten asymptotically to the magnitude of the cluster equivalent to using an infinite aperture. In practice this is not always the case, so the profile of each cluster was inspected visually to remove any sources that would pollute the final aperture corrections. This included sources with dips in the profiles, and those that had not flattened sufficiently, but were still rising at 15 pixels (likely due to the sampling of nearby sources). 

We were left with 8 clusters with reasonable profiles in all bands. The aperture corrections were calculated by averaging the difference between the 5 pixel and 15 pixel apertures in each band: 0.40, 0.34, 0.34, 0.32 and 0.36 magnitudes for the UV, U, B, V and I bands respectively. These corrections were subtracted from the magnitudes of each cluster. 

To further reduce likely erroneous photometry, cuts were then applied across the U, B and I bands in addition to the current cut in the V band. Any sources fainter than an apparent magnitude of 26 in all bands were removed. This ensured all final sources were reliably detected across these key wavelengths for the purposes of age and mass fitting carried out later. Detection in the UV band was useful if possible but not essential to the study. A final cut in the photometric errors was then applied across all sources. Objects with an error $\ge 0.2$ were removed. At this point, automated computer-based methods are poor at refining the catalogue and reducing the number of contaminants, so source inspection by eye was carried out on the remaining 3802 objects, as per \cite{ba12, silva14} and \cite{adamo15}. The refinement process involved inspection of each object's radial profile to ensure it was a single extended source (preferably mostly circular and symmetrical) and not multiple neighbouring stars within the same aperture or an association. Unresolved single stars could also be identified this way using the magnitude and sharpness of the peak in the profile. This process is not infallible, however, as not all clusters are symmetrical and may not be brightest in the centre. 

Each source was labelled according to the degree of likelihood of being a true cluster. 'Class 1' was assigned to sources that were likely clusters, i.e. the profiles matched a cluster profile and they were symmetrical and extended. 'Class 2' was assigned to sources that could potentially be clusters but were more likely to be associations. 'Class 3' was assigned to objects that were not clusters, such as unresolved point sources or possibly background galaxies and were later removed from the catalogue completely.

\subsection{Age and mass fitting}
\label{sec:fit}

The magnitudes for each cluster obtained from photometry were used to fit ages and masses for each of the clusters. The fitting procedure was done as per \cite{a10ab}, by comparison of each cluster's SED with simple stellar population Yggdrasil models \citep{zack11} using a Kroupa IMF \citep{kroupa01}. The models incorporate super solar metallicity and account for nebular and continuum emission. 

Traditionally, H$\alpha$ magnitudes are used to break the degeneracy between age and extinction at young ages (e.g. \citealt{whitmore10}) and give more accurate estimates of cluster ages. However, the H$\alpha$ image for NGC 1566 is only available in WFPC2 data, with smaller coverage of the galaxy, missing many of our sources. SSP models assume that contributions to the flux from ionised gas emission are present in young clusters, however it has been shown that clusters as young as 3-4 Myr have expelled their remaining gas \citep{ba14, me15}. The early removal of gas means that the H$\alpha$ emission is lower than model predictions, which likely gives an older age for the cluster than required. Additionally, the distributed nature of H$\alpha$ means that it can be difficult to identify whether emission is actually associated with the cluster or association itself. H$\alpha$ is also more susceptible to contamination from nearby sources.  

Due to these factors we opted to use UV in the fitting procedure instead of H$\alpha$, as this can also disentangle degeneracies at young ages. Furthermore, UV emission is associated with massive stars within clusters, and so is more reliably representative of the cluster. The potential for using UV to disentangle degeneracies in age dating of clusters is a key aspect of the LEGUS survey \citep{calzetti15}. 

 \section{Colour-colour plots}
\label{sec:ccplot}

\begin{figure}
\includegraphics[width = 8.5cm]{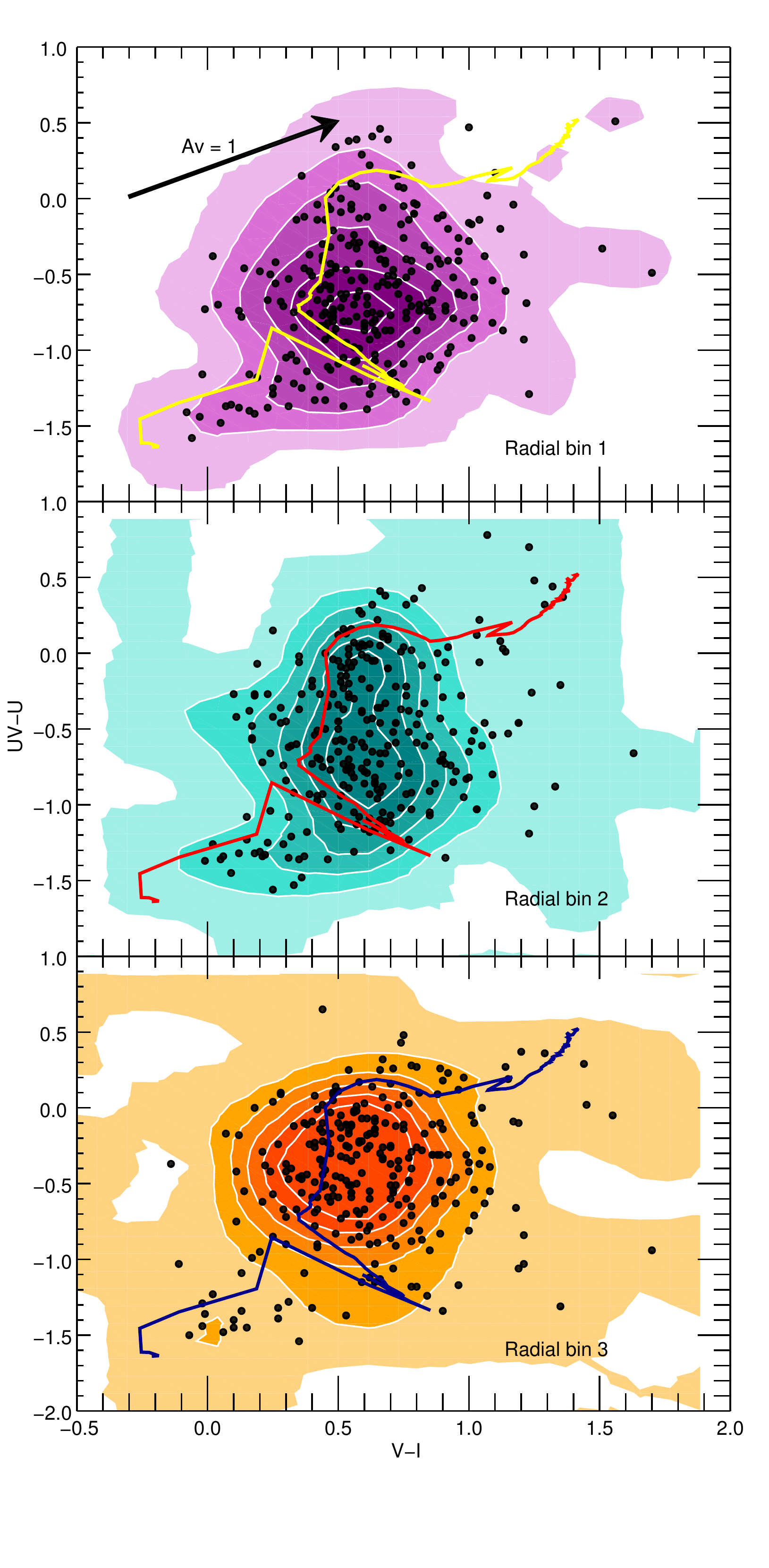}
\caption{Colour-colour plots for clusters in NGC 1566. The clusters have been split into equal numbers in radial bins of 0-3.3 kpc, 3.3-4.7 kpc and 4.7 kpc outwards. The plots show the traditional U-B vs V-I colours. They are overplotted with models showing the evolutionary tracks of clusters and contours are plotted to show the concentration of the clusters in colour space. The extinction vector showing the movement of clusters on the plot due to extinction effects is in the top left corner.}
\label{fig:ccplotub}
\end{figure}

An interesting result of previous studies of cluster populations has been that clusters closer to the central regions of the galaxy display different properties than those further towards the edge of the galaxy, showing variations in quantities such as age and mass. We expect outer cluster populations to contain more old clusters due to lower levels of disruption. 

A study of NGC 4041 found a marked difference in the position of clusters in colour space, with clusters closer to the centre of the galaxy in a bluer space than those in the outer regions \citep{kon13}. Additionally to NGC 4041, M83 also displays a variation in colour for clusters in outer and inner regions of the galaxy \citep{ba11}. In contrast to this finding, \cite{ryon14} report no difference in colour space for the cluster population of NGC 2997, though this study compared clusters specifically in the circumnuclear region to the rest of the disk, which could be considered slightly different to other studies.

Fig.~\ref{fig:ccplotub} shows our U-B vs V-I colour-colour plots for NGC 1566. These plots include a mass cut of 5000~\msun\ applied to the catalogue to account for stochastic sampling of the IMF and inaccuracies experienced in fitting low mass clusters, and includes only class 1 sources. The catalogue has been split into the three populations at different radial distances from the centre of the galaxy as described in \S~\ref{sec:test}, with each bin containing $\approx 270$ clusters. The evolutionary models for the clusters have been plotted over the points. 

The majority of the points lie in reasonable colour space with respect to the model track, meaning they can be traced back along the extinction vector to an age on the model. These clusters are to the right of the track.  

\begin{table}
\centering
\begin{tabular}{c c c}
\centering
 & Median U-B & Median V-I \\
\hline
Bin 1 & -0.90/ \textbf{-0.64}/-0.33 & 0.46/ \textbf{0.60}/0.78 \\
Bin 2 & -0.92/ \textbf{-0.57}/-0.18 & 0.46/ \textbf{0.60}/0.77 \\
Bin 3 & -0.68/ \textbf{-0.38}/-0.11 & 0.44/ \textbf{0.59}/0.78 \\
\end{tabular}
\caption{The difference in colour properties for cluster populations at different galacto-centric distances. The three bins are those used throughout the analysis. The median U-B and V-I colours (shown in bold) were calculated over the entire populations and then compared to see if there were major differences in the colour properties of the clusters at further distances from the centre. Values to the left of the median are the lower quartiles and to the right are the upper quartiles. There is little difference in the values for V-I, but U-B shows more variation. The second bin with two maxima on the contour diagram has the largest inter-quartile range, which would be expected of a more spread distribution.}
\label{tab:averages}

\end{table}

To investigate the distributions, we found the median U-B and V-I value for each population and found that they were all located in approximately the centre of the distribution, as shown in Table~\ref{tab:averages}. The observed trend is very similar to that found for M 83 \citep{ba11}. There is no variation in V-I colour, however there is clearly a small difference in the U-B colours, with the middle population displaying 2 peaks in the density of points in the centre, with each corresponding approximately to the values of the single peaks of the other bins. This indicates that the median U-B colour is continuously changing between bins. Clusters further towards the centre of the galaxy are slightly bluer, in agreement with the results for NGC 4041. The difference indicates that there could be a small variations in the ages of the clusters radially from the centre, as age would be the primary contributor to a change in colour. Potentially, these variations in age are due to the changing levels of disruption throughout the galaxy - with the highest near the galactic centre meaning that fewer older clusters are present.

\section{The luminosity function}
\label{sec:lumfunc}

\begin{figure}
\includegraphics[width = 8.5cm]{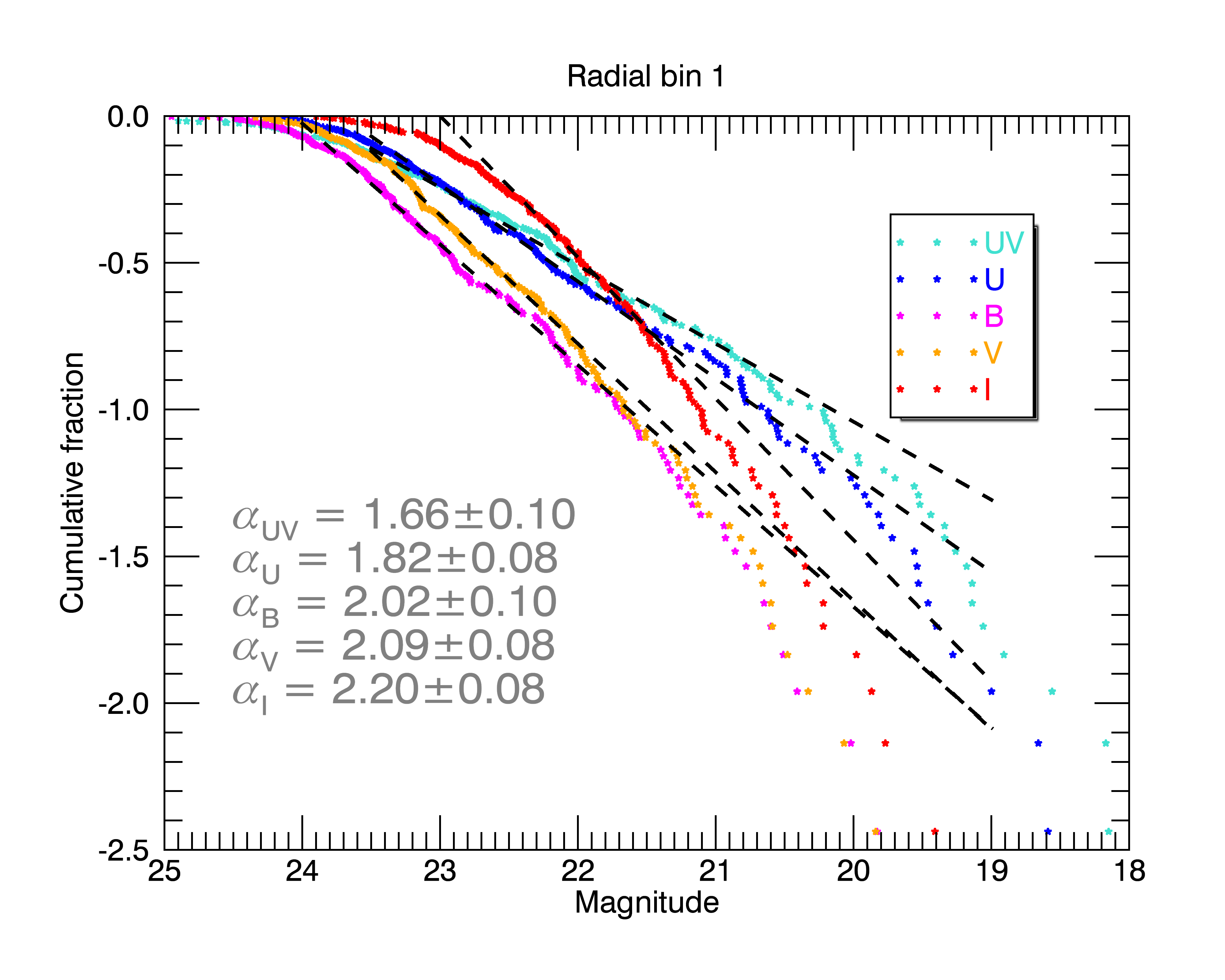}
\includegraphics[width = 8.5cm]{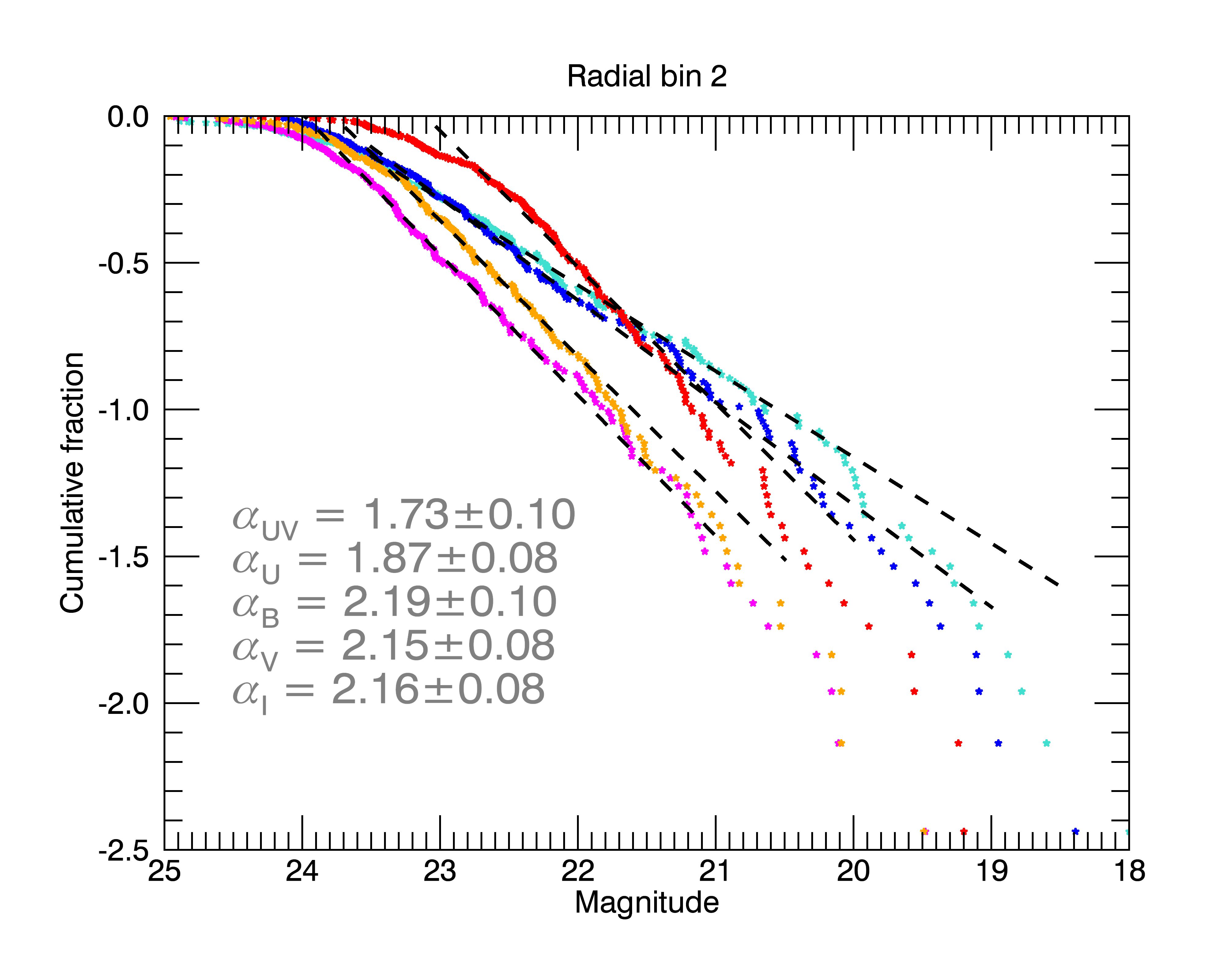}
\includegraphics[width = 8.5cm]{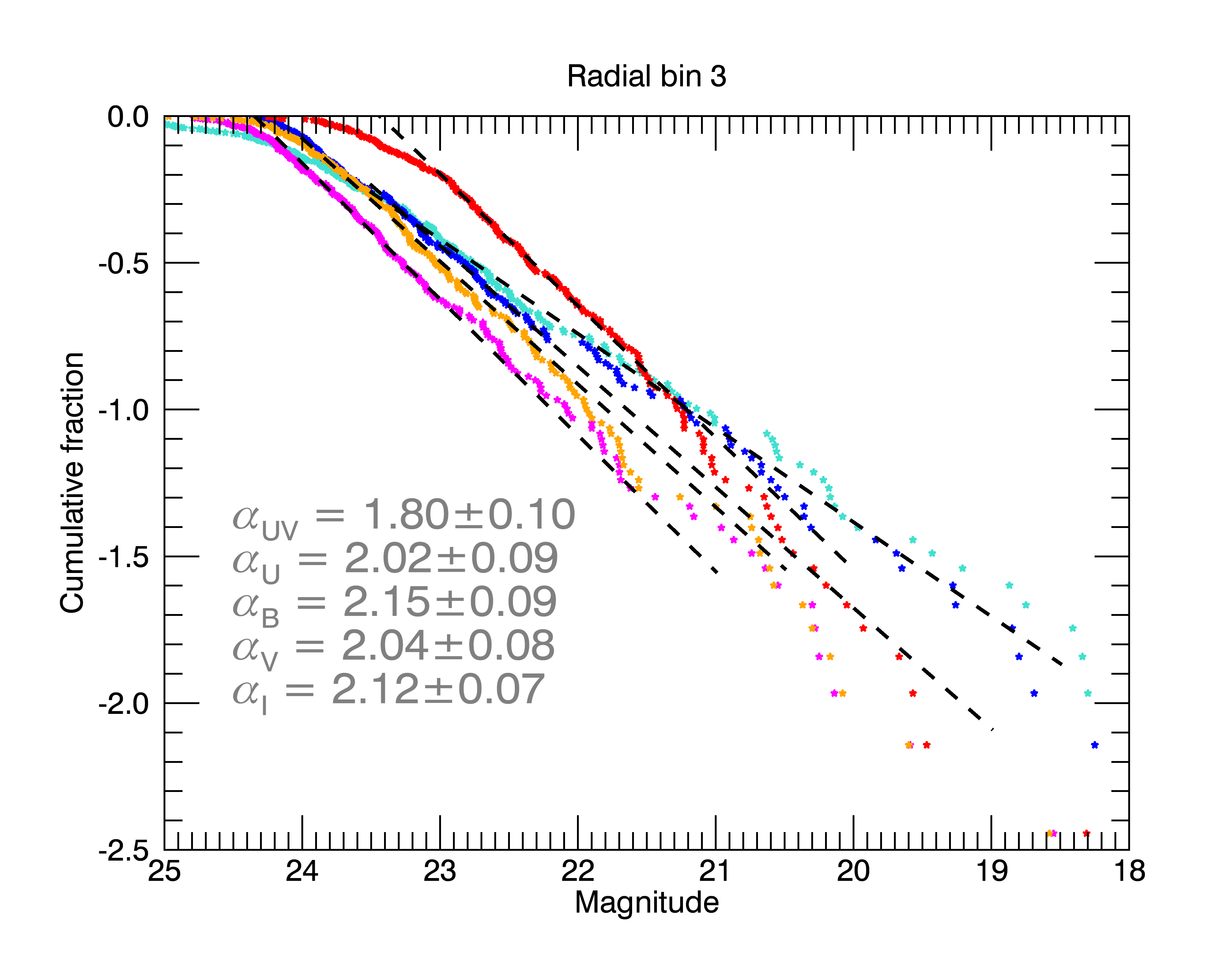}
\caption{Luminosity functions of the three binned regions of the galaxy as a log plot of the cumulative fraction, shown for each band. The shape of the function can be approximated by a power law, as we would expect, and has been observed for other galaxies with similar studies, such as M83. The values displayed on the plots are the gradient of the fit to the power law section of the function, using the \textsc{linfit} line-fitting utility in \textsc{idl} with errors found from fitting synthetic populations created by Monte Carlo techniques.}
\label{fig:lumfunc} 
\end{figure}

\begin{figure}
\includegraphics[width=8.5cm]{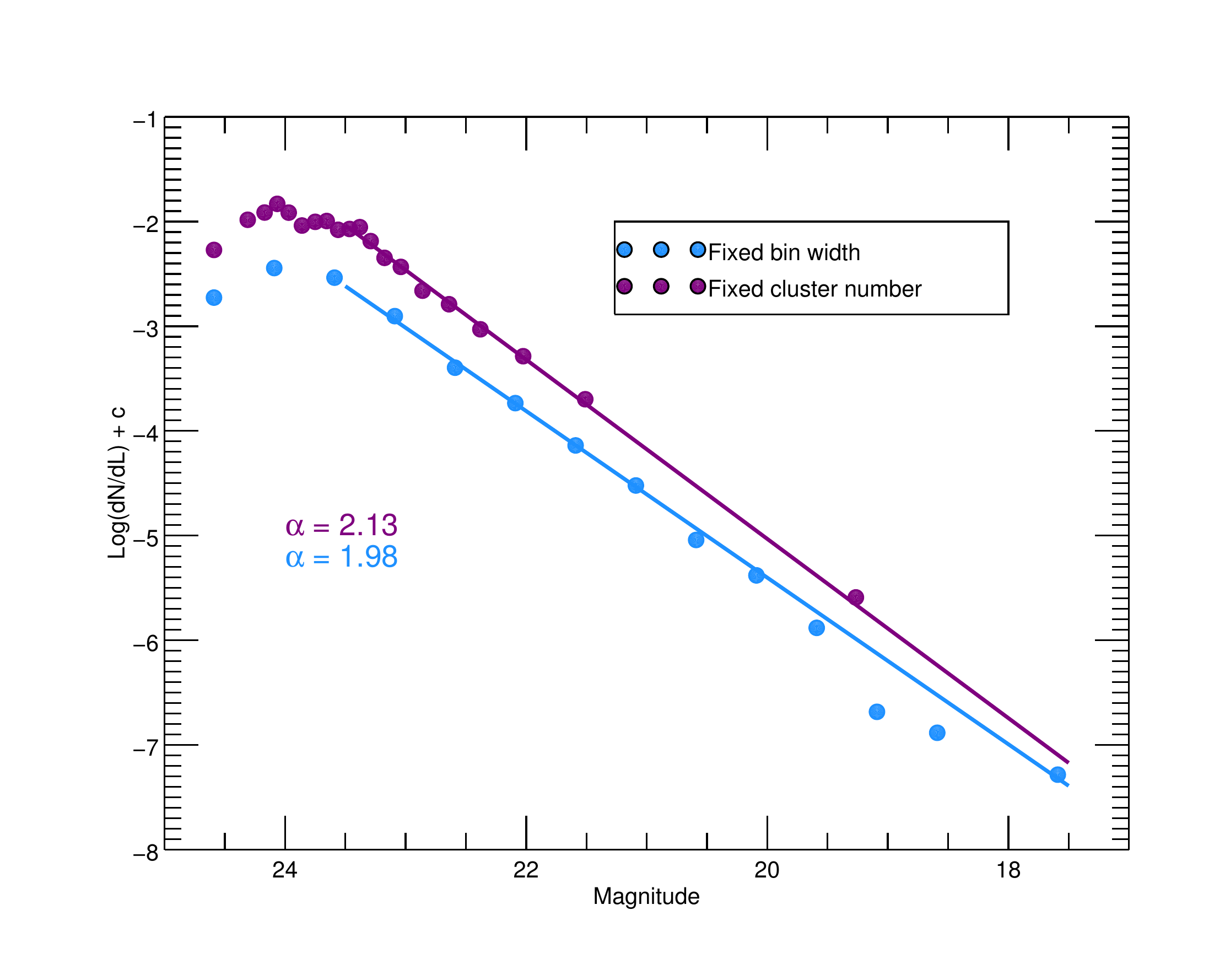}
\caption{The luminosity functions for the whole cluster population above $5000~\msun$ for class 1 sources in the B band created by binning the clusters, rather than using a cumulative function. The purple points show a binned function using variable bin width with equal numbers of clusters in each bin and he blue points represent a function with fixed bin width and variable numbers of clusters. The turnover at low luminosity is due to incompleteness and occurs at approximately the same values as for our cumulative function. Far less of a truncation at the bright end is seen in the binned functions, which can easily be dismissed as being within Poisson noise, when in fact this could be a physical effect.}
\label{fig:binlf}
\end{figure}

Many previous cluster population studies have explored the luminosity function of  clusters in other galaxies. As luminosity is proportional to mass (modulo age), this can provide insight into the behaviour of the cluster mass function using an observed quantity, rather than a modelled one. While this section explores our observed luminosity functions, in \S~\ref{sec:predictions} we model the function using key values from our cluster population and compare to our data.

Fig.~\ref{fig:lumfunc} shows the luminosity functions for the three radial regions selected, showing how the luminosity function for each band varies with distance from the galactic centre. The plots show separate luminosity functions for each of the UV, U, B, V and I bands using the log of the cumulative fraction for the clusters plotted against their magnitude.

The cluster luminosity function is believed to behave largely as $N dL \sim L^{-\alpha} dL$, though the shape is not usually a pure power law, but one that is truncated at the brighter end and can sometimes be better approximated with a double power law function (e.g. \citealt{gieles06a}). {Fig.~\ref{fig:lumfunc} shows a levelling at the faintest end, likely} due to a combination of reaching the reliable detection limit and potentially the disruption process, which primarily affects lower mass clusters if a MDD disruption regime is considered. The truncation at the brighter end is caused by a truncation in the cluster mass function at high masses, though this is not always in a one-to-one ratio as differential fading across wavelengths alters the position of clusters of the same mass in the luminosity function. We find the gradient of the power law section ($\alpha$) of the curve by using the \textsc{linfit} utility in \textsc{idl}, which minimises $\chi^2$. 

The range of luminosities selected for the fits varied slightly between each band, as the position of the power law section changed. The flat end of the distribution at faint luminosities was never included in the fit as the cause of the shape is most likely dominated by our detection limits and magnitude cuts, with little influence from physical effects. Additionally, the brighter end of the distributions were also not included due to the low number of clusters and therefore poor statistical significance. The ranges of fits were within 21-24 magnitudes across each band and each bin; Table~\ref{tab:fits} shows the exact fit ranges for the first bin, with other bins being approximately the same values. The fits are plotted over the curves for each band, and the values of $\alpha$ for each fit are displayed on the plot.

\begin{table}
\centering
\begin{tabular}{c c c}
Band & Min fit & Max fit \\
\hline
UV & 23.5 & 21 \\
U & 23 & 21.5 \\
B & 23.5 & 22 \\
V & 23.5 & 22 \\
I & 22.5 & 21.5 \\
\end{tabular}
\caption{The maximum and minimum magnitudes used for the fit of each of the lines in the first radial bin. Other bins were fitted with approximately the same values.}
\label{tab:fits}
\end{table}

Other similar studies may use a binned luminosity function, as shown in Fig.~\ref{fig:binlf} for the B band. All class 1 clusters above our mass cut of $5000\msun$ are used and we show the results for variable bin width (purple) and fixed bin width (blue). The turnover at the faint end is due to incompleteness and occurs at approximately the same magnitude as the equivalent turnover in Fig.~\ref{fig:lumfunc}. There is less evidence of a truncation at the bright end as seen in the cumulative functions, however the dramatic increase in bin size (for variable bin width) in the bright bins containing the same number of clusters indicates that there are far fewer bright clusters and that there could potentially be a truncation, which is not clearly evident from a binned function. The slight dip in the fixed bin width distribution could be easily dismissed as Poisson noise, when in fact the effect could be physical. A cumulative function is more sensitive to changes at the bright end, though with the caveat of low cluster numbers and therefore poorer statistics.

Previous studies have found that the value of $\alpha$ is usually $\approx 2$, with small variations \citep{degrijs03}. Studies of M83 show that the value is slightly higher for clusters in the outer regions of the galaxy, indicating a decrease in the number of bright clusters \citep{ba12}. The inner areas of galaxies (just outside of the bulge) would be expected to experience higher levels of star formation than further out into the arms due to a higher density of molecular gas. \cite{ryon14} find slightly different results for NGC 2997. The circumnuclear regions were found to be slightly shallower than the disk in the U and B bands but steeper in the V and I bands.

Our results agree with the other cluster population studies; $\alpha \approx 2$ with small variations. There are only very small variations in the slopes for clusters between radial bins, with the most prominent differences seen in the UV and U bands. These variations are within the errors on the fit and therefore no definite trend can be determined from these values. 

A variation has also been seen in the index of the power law slope between bands of different wavelength. The bluer UV and U bands consistently have shallower slopes than the redder bands in several galaxies \citep{ryon14, ba14, gd12}. The reason for this trend is likely differential fading of clusters across the different filters. Bluer bands fade more quickly than redder bands so clusters of the same mass will contribute across a wider luminosity range in the UV, for example, while occupying a smaller range in the I band. This spreading out of clusters in the UV and U bands creates a shallower slope \citep[e.g.][]{gieles06a}.        

The effect of wavelength is observed, with bluer bands generally having shallower slopes by up to 0.54, larger than the fit errors. This could suggest a lack of disruption in this galaxy, alternatively, this effect is more likely due to the rapid fading of clusters in bluer bands compared to red, as mentioned previously.    

The luminosity function has been observed to be consistent with indices of 2 for associations in addition to tightly bound clusters \citep{ba07}. To investigate this effect within NGC 1566 we plotted the same luminosity functions but only for class 2 sources (i.e. associations and groups). We found that the fits were generally steeper than for class 1 sources, with all between 2 and 3. The slight difference could potentially be explained by the quality of the sources. Unlike class 1 sources, the objects defined as class 2s are more likely to be incorrectly identified as associations, and the sample is more likely to be contaminated with stars \citep{whitmore14}. The sample could potentially be polluted with many unreliable sources with poor photometry. Despite this, the value is not drastically different and, in agreement with other studies, points to a process of star formation independent of the scale of the ISM \citep{elmegreen02}.     

\section{Age and mass distributions}
\label{sec:agemass}

\begin{figure*}
\centering
\includegraphics[width = 17cm]{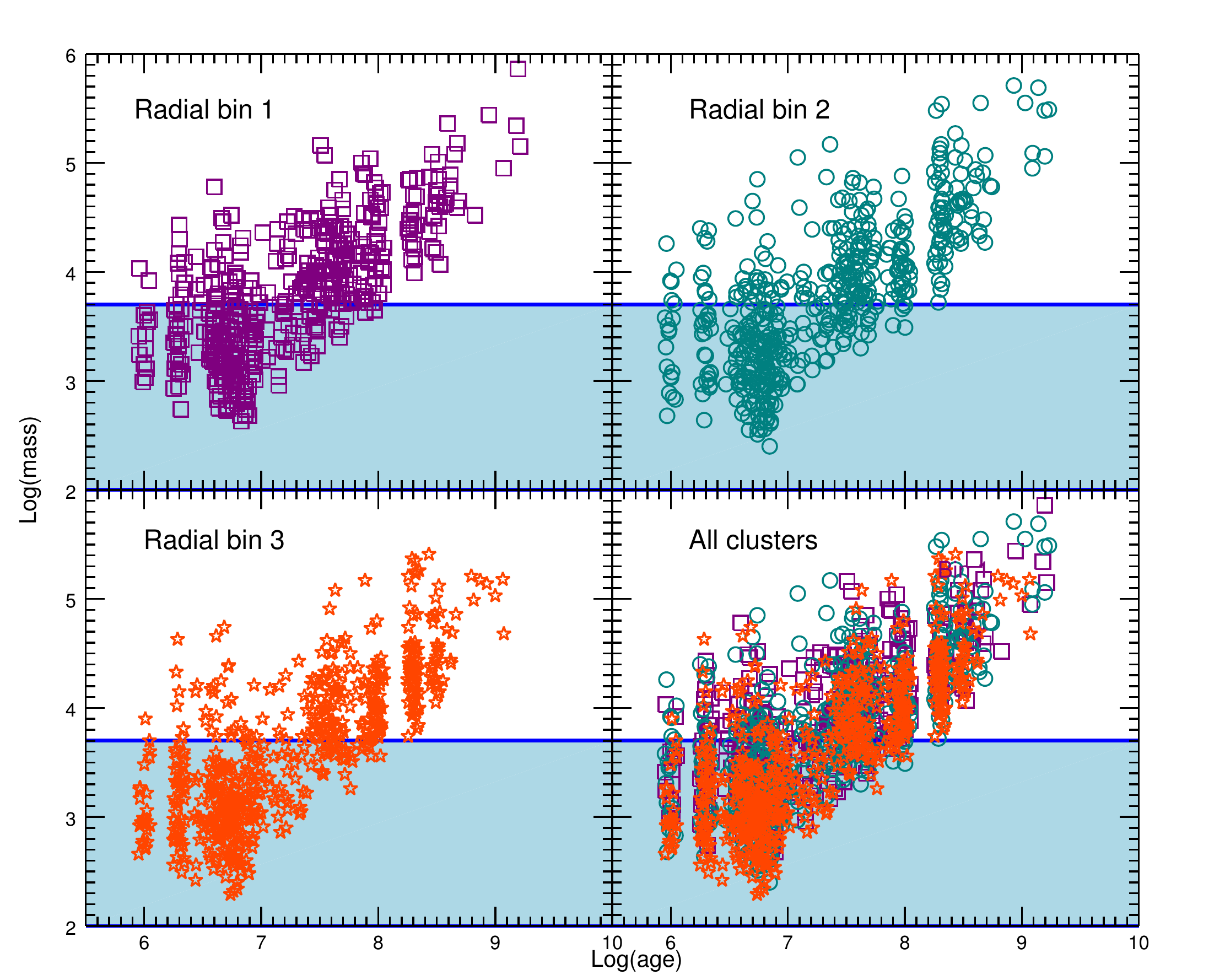}
\caption{Age vs mass for all class 1 sources in the catalogue. Each bin is displayed separately, with the combined population in the final plot. The blue line indicates the age cut we apply to the catalogue for the other analyses. We lose $\sim$ half of all clusters by applying the mass cut. The lower limit line on the right side of the plots represent the detection limits in the catalogue, as well as the magnitude cut imposed during the catalogue creation. The apparent ceiling to the data is due to the mass function.}
\label{fig:agemass}
\end{figure*}

The mass and age distributions of a population of clusters are highly useful for studying the star formation history of the galaxy and the effects of the disruption process. Cluster population studies (including this one) generally impose a mass cut on the population. This accounts for inaccuracies in age and mass fitting and stochastic sampling of the stellar IMF \citep{silva11, foues10}. Additionally, a mass-limited sample prevents bias in the age distribution caused by young clusters. Fig.~\ref{fig:agemass} is the age-mass diagram for class 1 sources in our catalogue. There appears to be very little difference between the three populations of clusters in varying distances from the galactic centre. The blue line indicates the mass cut we applied to the catalogue for our analyses at $5000~\msun$. We lose a large percentage of clusters after applying the cut, ($\sim50\%$), however this step is necessary for the reasons mentioned previously.

The limit that appears above the data is due to the mass function - clusters cannot form to an infinitely large mass - the maximum and likelihood is determined by the mass function. Additionally, the cut off of data on the lower right hand side of the plot is due to the detection limit of the data. A magnitude limit is imposed by our detectors, which creates the sharp cut off of visible clusters, and which is used to determine the age and mass at which populations are incomplete - our sample appears to be incomplete after $\approx 100-200$ Myr, when lower mass clusters become too faint to observe.

The age-mass diagram also displays some less populated areas, such as the gap around 10-30 Myr. This is a well-known artefact from the mass and age fitting process and has been previously identified in other galaxies such as M51 \citep{bik03, nb05} and M83 \citep{chand10}.

\subsection{Number of clusters per age bin}
\label{sec:agedist}

\begin{figure}
\includegraphics[width=8.5cm]{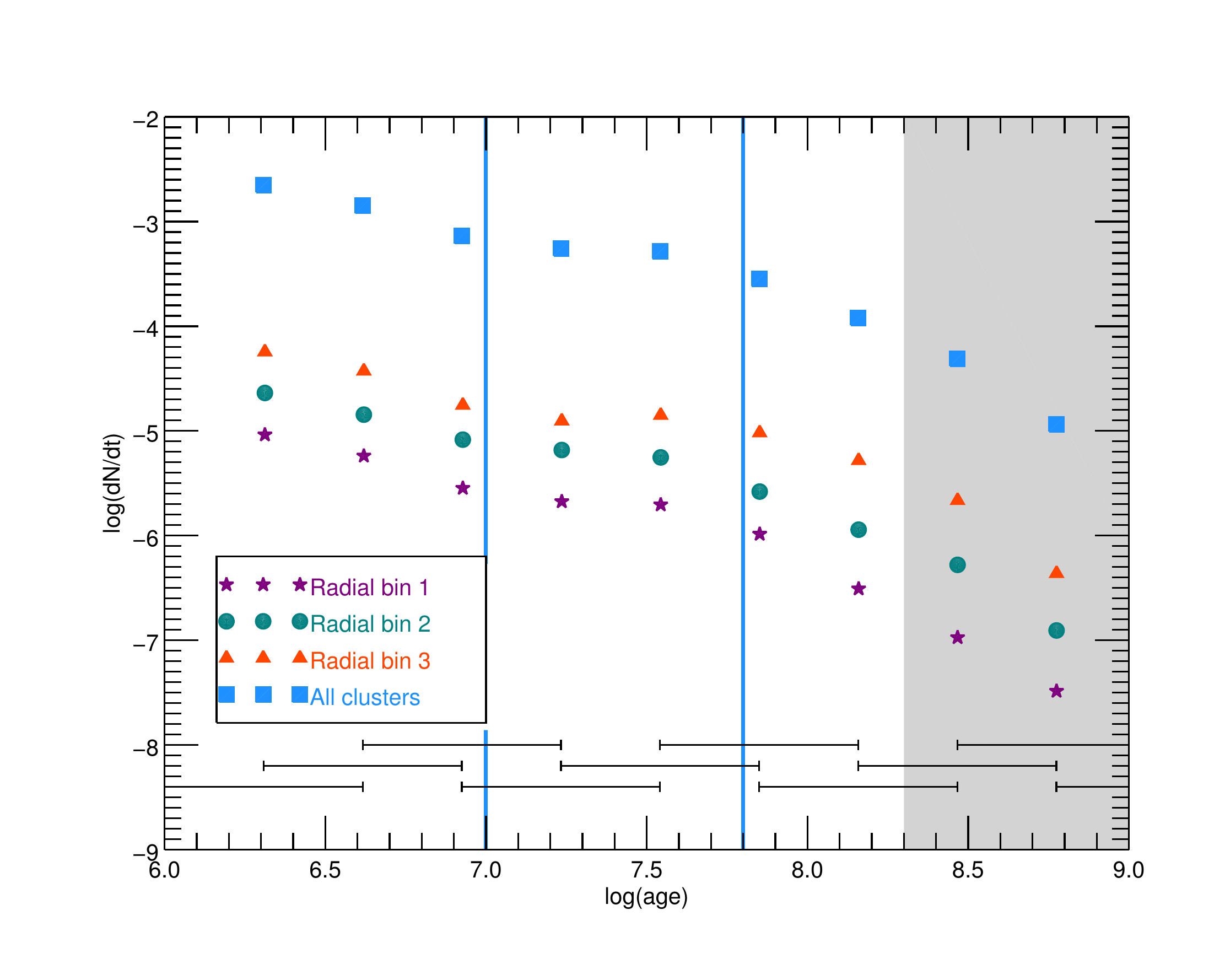}
\caption{The age distributions for different sections of the galaxy. Only class 1 sources above 5000 $\msun$ have been included. Overlapping bins have been used to remove unphysical variations caused by the binning procedure, with the coverage of each bin shown by the bars at the bottom of the plot. As shown, there is little difference between the shapes of the distributions. A factor has been added to each of the lines to separate the points and make the shape of the distribution easier to see. The two blue vertical lines indicate the separate regimes within the age distribution, as described in the text. The plot is shaded after 200 Myr as we are incomplete here, and may be partially incomplete from 100 Myr.}
\label{fig:agedist}
\end{figure}

Age distributions of clusters across the galaxy can provide information on the formation history of the clusters and the scale or strength of the effect of disruption. By studying different areas of the galaxy you can also determine whether the disruption process is environmentally dependent. It can be assumed that the shape of the age distribution is only dependent on cluster formation and disruption as our sample is mass-limited. This means that a minimum detection luminosity should not be shaping the distribution until after 100 Myr (e.g. \citealt{gieles07}), as can be seen in the age-mass diagram.

Fig.~\ref{fig:agedist} shows the age distributions of clusters in the three radial subsections of the galaxy, separated by distances of 0.5 in log space to clearly show their shape. The blue squares show the shape for the entire cluster population, while the purple, teal and red points show the consecutive radial bins. We use overlapping bins to minimise the effects of the stochastic nature of bin fitting, with the bin sizes shown at the bottom of the plot.

The shape of an age distribution can generally be roughly approximated by a single power law with steepening at high ages due to incompleteness. Our plot displays potential evidence of a three component shape predicted by a system of mass dependent disruption, as summarised by \cite{lamers09}. The first section ranges from the beginning of the plot to $\sim 10$ Myr. This decline is caused by the dissolution of young clusters as they expel any gas left over from the star formation process, and should be independent of mass \citep{ba06}. However, at least part of this drop is likely due to the inclusion of unbound associations in our cluster sample \citep{ba12}.  

\begin{figure}
\includegraphics[width=8.5cm]{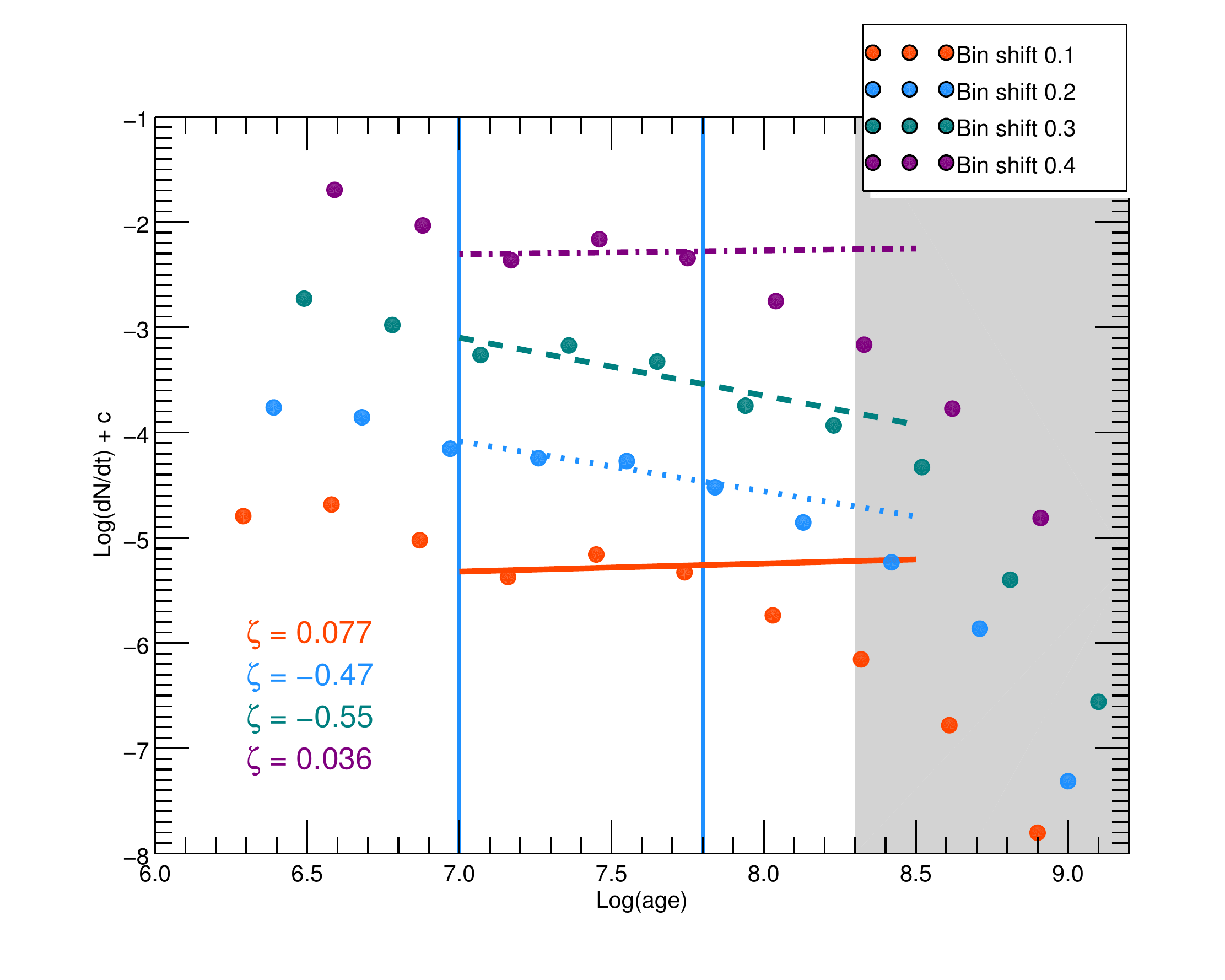}
\includegraphics[width=8.5cm]{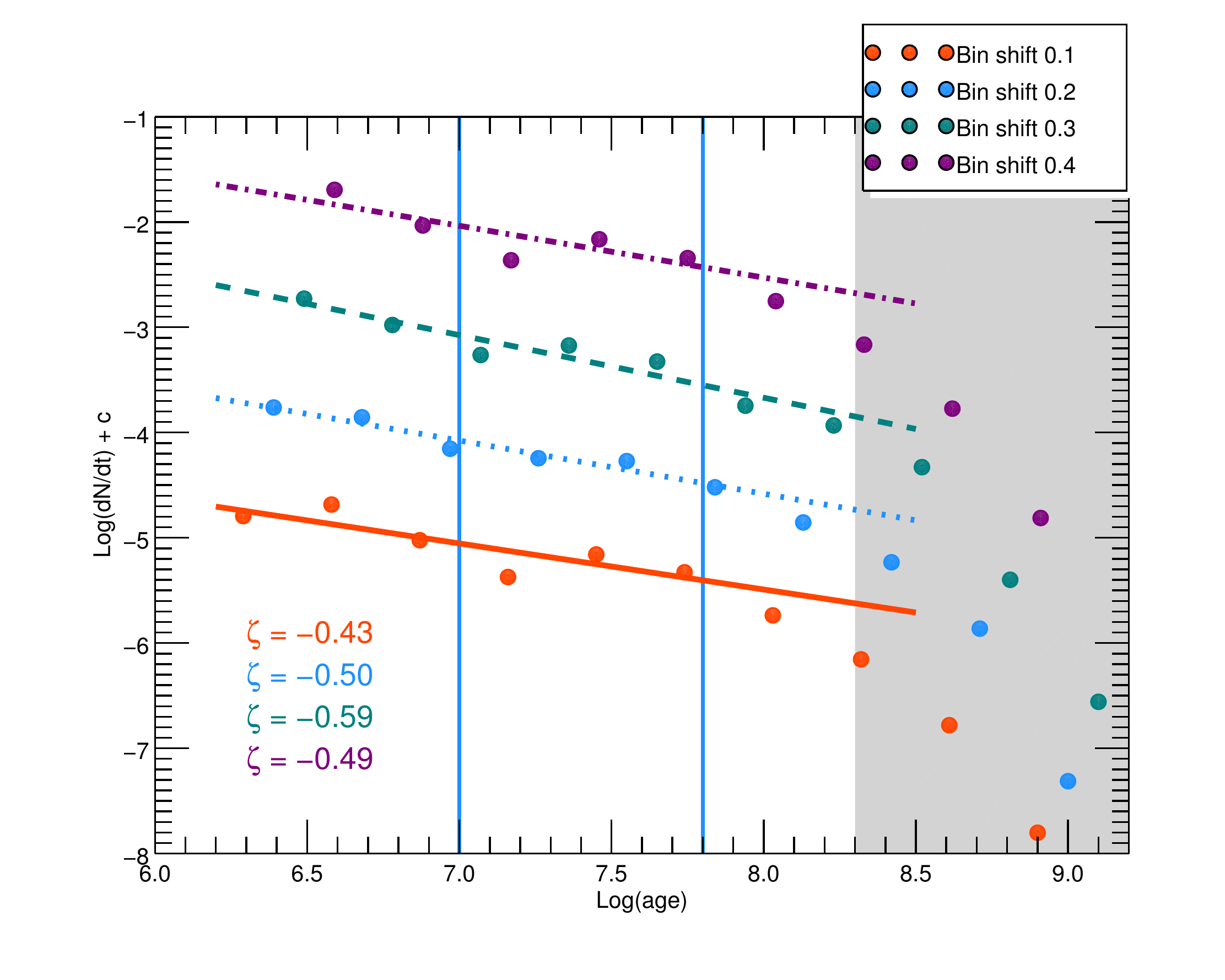}
\caption{The effect of shifting bins in the age distribution for all clusters more massive than $5000~\msun$ and only class 1 sources. The shape still displays evidence of a flattening in the 10-100 Myr range so fitting single power laws to the data can be misleading. For this reason we do not attempt to fit our age distributions. This plot uses the same binning procedure as our age distribution. The top plot is fitted from 10-100 Myr, not including clusters $<10$ Myr, as per \protect\cite{chand14} and the bottom plot includes all clusters below 100 Myr, as per \protect\cite{silva14}. }
\label{fig:agecheck}
\end{figure}

As our data is mass-limited, the next section, up to $\sim 100$ Myr for NGC 1566 (though this can extend to much more advanced ages in other galaxies), is fairly flat, which indicates little disruption. The age-dating artefact at 10-30 Myr could be affecting the shape of the distribution in this section, however we think this is unlikely, as clusters in this age range will still be accounted for in nearby bins, and shifting bins still produces a flat section (see Fig.~\ref{fig:agecheck}). The final section is a strong decrease again, which could potentially be due to mass-dependent disruption of clusters due to tidal effects with a large contribution from incompleteness of the fainter, older clusters \citep{bout03}.  As we are incomplete after 200 Myr (and potentially partially incomplete from 100 Myr) these two mechanisms cannot be entirely disentangled. Approximate values of the changes in the sections are shown as vertical lines on the plot, and the shading indicates the age regime in which we are incomplete.

The age distribution can be highly susceptible to changes in the binning procedure used to represent the data, though binning is important to overcome small variations caused by artefacts in age fitting. Fig.~\ref{fig:agecheck} shows how shifting bins can alter the age distribution. Each shift still displays evidence of a flattening in the $\approx 10-100$ Myr range. The top plot fits from 10-100 Myr as done in \citep{chand14}, while the bottom plot includes all clusters below 100 Myr, as per \citep{silva14}. As the fit range is so small in the top plot, the fit varies wildly between -0.55 and 0.077. The bottom plot is more consistent in fits and ranges from -0.59 to -0.43. The multi-component behaviour of the age distribution shows that a single power law is not a good fit to the data, and could be misleading in either case.

Studies have shown that the age \dist\ is strongly related to the star formation history (SFH) in the galaxy (e.g. \citealt{bastian09}). The distribution can be considered to be independent of cluster formation history if the galaxy is known to have a quiescent history with a constant low star formation rate. It is unknown if this is the case for NGC 1566, though the current fairly high rate of star formation (4.3 $\msunyr$ \citep{thilker07} (amended, discussed later in \S~\ref{sec:gamma})) is likely not a new development as there does not appear to be an over-abundance of very young clusters. This could suggest a star formation history with a fairly high constant rate of formation.  Additionally, while most studies can assume a constant star formation history due to little activity in the galaxy, NGC 1566 is a Seyfert galaxy and is a member of a galaxy group, potentially affecting its recent star forming activity.

A theory of disruption independent of environment would be expected to have the same shape of the age function in all areas of the galaxy as clusters should disrupt uniformly, unaffected by their environment. In this scenario, a single power-law age distribution is expected, with an index of $\sim-0.8 - -1$. (e.g. \citealt{whitmore07}). More recently, \cite{chand14} have claimed that variations in fit between fields in M 83 is due to differences in the SFH between fields, rather than varying levels of disruption. NGC 1566 displays the same shaped distribution for each bin and an overall shape potentially indicative of a disruption mechanism dependent on mass. The flat section seen between $10$ and $100$~Myr argues strongly against a quasi-universal age distribution.  However, if a single power-law is fit to the data over the full range a relatively steep index is found, due to incompleteness at high ages, and the abundance of young clusters in our sample at ages less than $10$~Myr. 

Another reason for the steepening of the age distribution at young ages could be observational biases. NGC 1566 is a fairly distant galaxy at $\approx 17$ Mpc, and therefore the densest clusters in the galaxy may appear as point sources, which would be removed by the concentration index cut or visual inspection phases. If the age distribution is biased towards larger clusters or associations that appear as clusters, the age distribution would decline more quickly due to the shorter lifetimes of less dense systems.

\subsection{Observed cluster mass function}
\label{sec:massfunc}

\begin{figure}
\includegraphics[width=8.5cm]{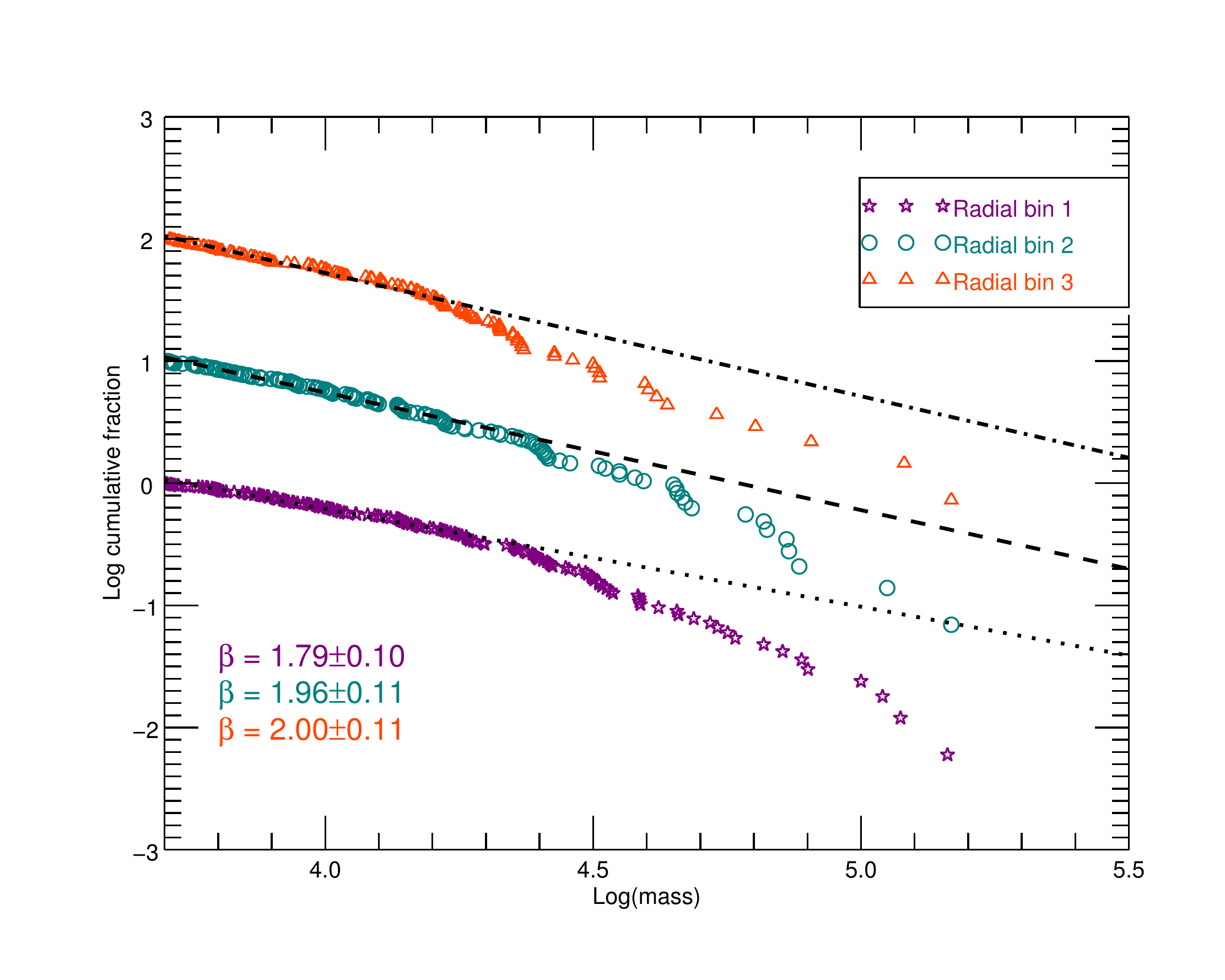}
\caption{The mass distribution for clusters in bins of distance from the centre of the galaxy. The fit to distribution is highly dependent on the range over which the line is fitted. Clusters older than $\approx 100$ Myr were removed as we are incomplete at high ages. Clusters younger than 10 Myr have also been removed due to possible incompleteness and difficulty in age and mass fitting in this age regime.}
\label{fig:massdist}
\end{figure}

The mass distribution of clusters can be described by $N dm \sim M^{-\beta} dm$; a function well approximated by a power law with a possible truncation at the high mass end. Fig.~\ref{fig:massdist} shows the mass distributions for the three radial bins for clusters older than 10 Myr and younger than $\approx 100$ Myr. This age range was selected as we are  likely partially incomplete after 100 Myr (as shown by the age-mass diagram in Fig.~\ref{fig:agemass}) and potentially affected by contamination from associations in the sample below 10 Myr (e.g. \citealt{ba12}). Additionally age and mass fitting is less accurate below 10 Myr. Only class 1 sources and those more massive than 5000~\msun\ are included. The distributions were fitted by minimising $\chi^2$ from $\approx 5000-30000~\msun$. There is a variation seen in the best fit for the power law section with respect to galactic radius, with a steeper slope for the outer clusters than for the inner two bins, potentially indicating the presence of more lower mass clusters than in the inner regions, which would agree with an environmentally dependent process of disruption. However, the differences are within the estimated errors on the fits and therefore no definite trend can be determined. 

\begin{figure}
\includegraphics[width=8.5cm]{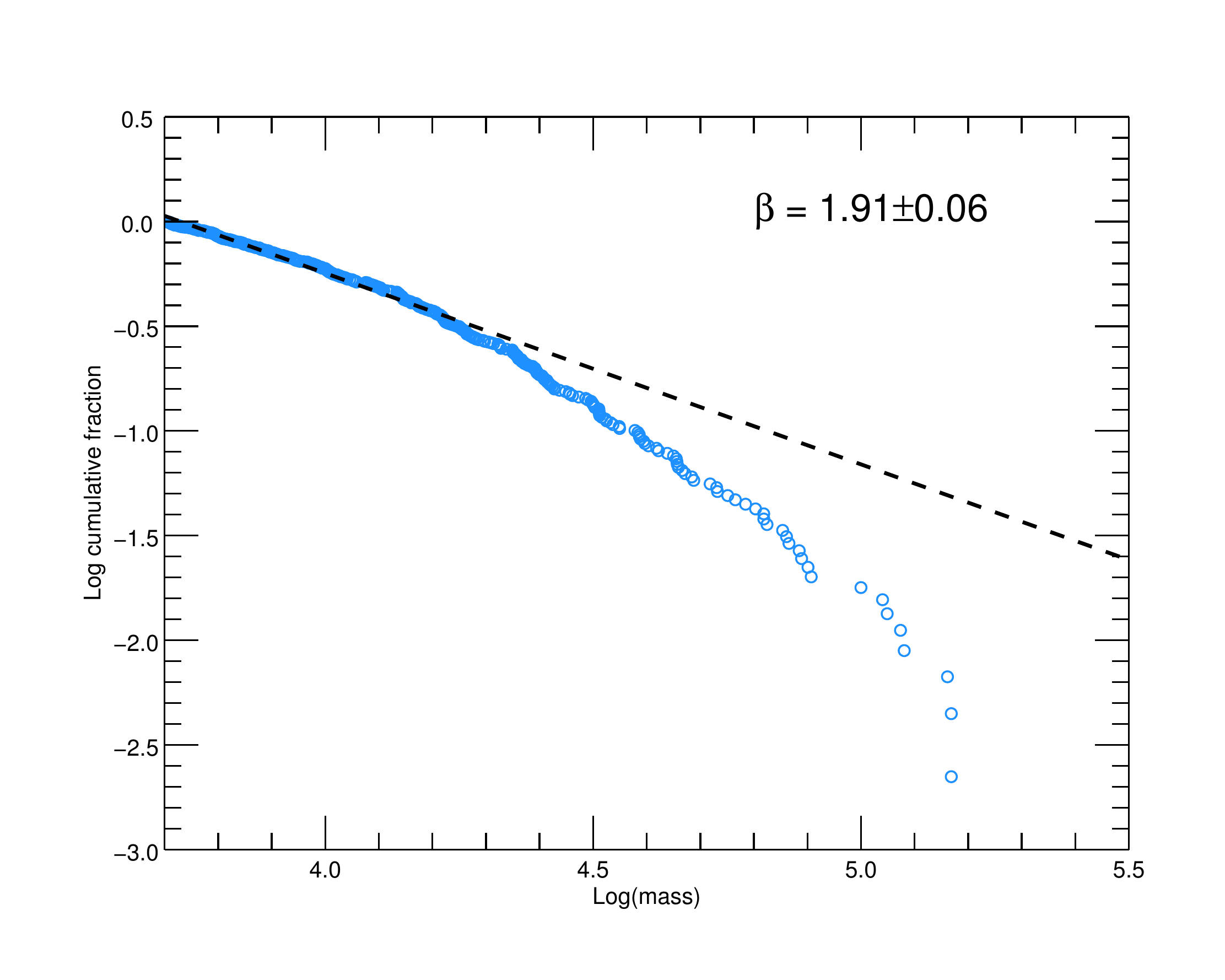}
\caption{The mass distribution for all class 1 sources in the catalogue, with a mass cut of 3.7 in log space. The best fit to the power law section is shown in the top right.}
\label{fig:massdistall}
\end{figure}

Fig.~\ref{fig:massdistall} shows the distribution for all clusters younger than 100 Myr and older than 10 Myr. The high mass end of the distribution does display evidence of a truncation as it deviates from the power law fit line (shown as black dashed line), but as this could be due to dwindling numbers of clusters at these masses. This effect is unlikely to be due to disruption in a MDD regime, as more massive clusters are affected less by disruption but it could play a part if MID is considered. The errors on the fits are found through Monte Carlo simulations of mass distributions.

\section{Comparison of NGC 1566 with modelled quantities}
\label{sec:models}

While much information can be gleaned from studying the observed distributions of quantities such as age, mass or luminosity for the clusters, the underlying processes causing their appearance are not always evident solely from observations. The modelling of physical processes underpinning the evolution of clusters is important in relating observables to the physics behind them. In this section we fit data to obtain information about the cluster population and then use this to form model functions to which we can compare our data and further understand its qualities. 

\subsection{Cluster disruption in NGC 1566}
\label{sec:disruption1566}

\begin{figure*}
	\centering
	\includegraphics[width=14cm]{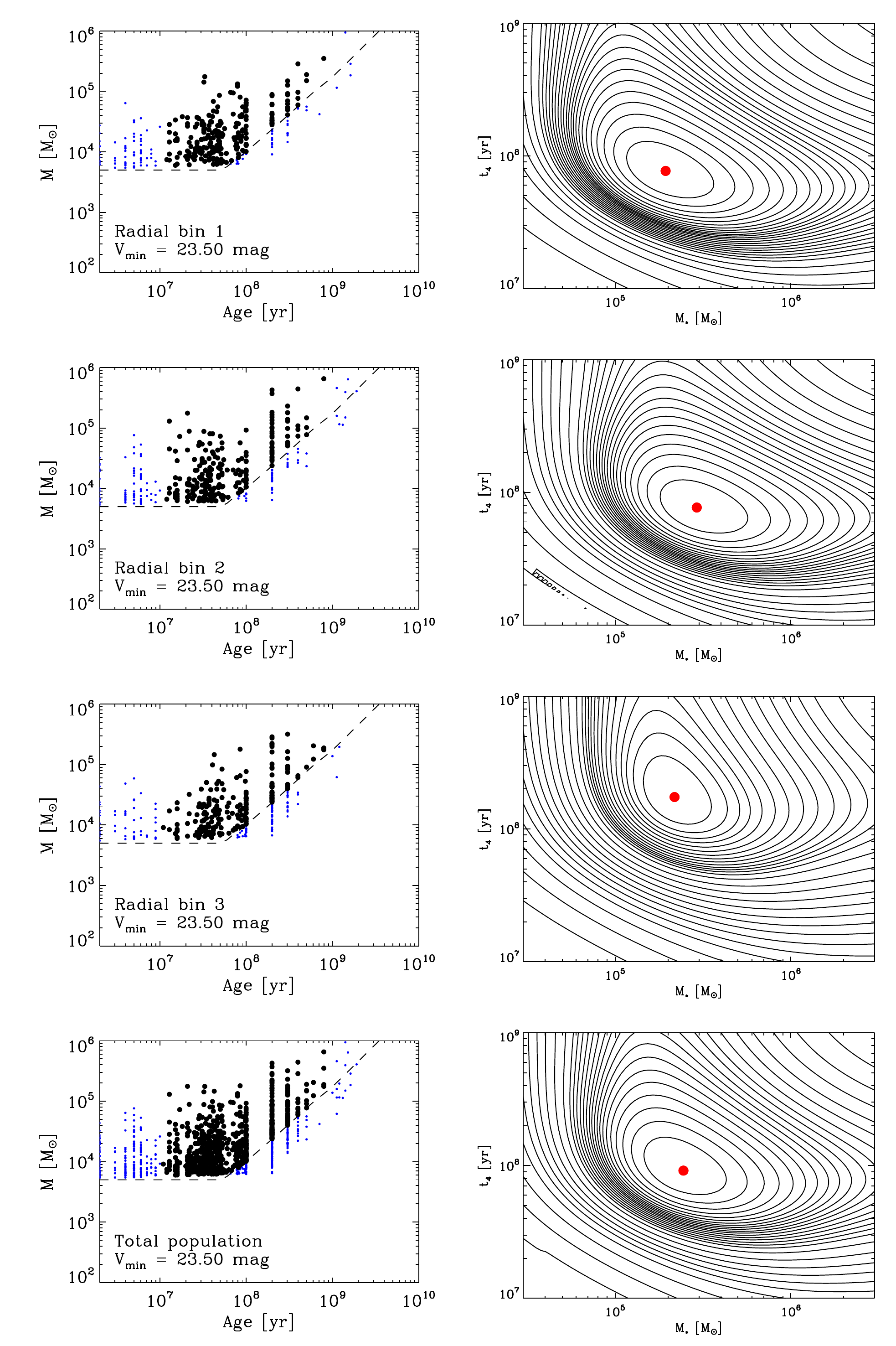}
	\caption{Maximum likelihood fits of the age mass distributions for the three different radial bins used throughout this study, and additionally the fit for the entire cluster population in the lowest plots. The left column shows the age mass plots of the clusters, while the right plots show the maximum likelihood fit for the truncation value of the mass function ($M_c$) and the corresponding timescale for disruption of a $10^4~\msun$ cluster. The clusters in black are used for the fit and are selected using an age cut of 10 Myr and a magnitude cut in the V band of 23.5.}
	\label{fig:maxlike} 
\end{figure*}

Age and mass distributions can provide empirical reasoning for the strength and scale of disruption in NGC 1566, however further information requires modelling and fitting data. Fig.~\ref{fig:maxlike} displays maximum likelihood fits to the age and mass plots for the three radial bins and the total population, as done for M 83 by \cite{ba12}. Clusters older than 10 Myr are selected for the fits, with an additional constraint of a minimum V band magnitude of 23.5 magnitudes, to ensure completeness. 

The fits employ the relationship between disruption timescale and cluster mass, scaling as $M^\gamma$ with $\gamma = 0.65$ and calculate $M_c$ and $t_4$ (the turnover mass and average time scale of disruption for a $10^4~\msun$ cluster respectively). We estimate $M_c$ $\approx 2.5\times10^5 \msun$ and $t_4$ as $\approx 100$ Myr for NGC 1566, taken from the fit for the entire cluster population.

There is little difference in $M_c$ for each successive radial bin. This suggests the truncation value in the mass function for each section should occur at the same mass. The truncation value in Fig.~\ref{fig:massdist} is difficult to determine accurately for the three populations due to dwindling cluster numbers at high masses. $t_4$ is very similar for the inner two radial bins, however is around a factor of 2 larger in the outermost bin. This result may not be significant but indicates that outer clusters are possibly disrupted more slowly than inner clusters. 

The values for $M_c$ and $t_4$ were used to model the contributions from different aged stellar populations to the luminosity function for NGC 1566, as discussed in the next section. $t_4$ was scaled to $t_0$ (the average disruption timescale of a 1~\msun\ cluster), as required for the model. $t_4$ was found to be 100-200 Myr, for which we took the average of the two values.   

\subsection{Luminosity function modelling}
\label{sec:predictions}

\begin{figure}
\includegraphics[width=8.5cm]{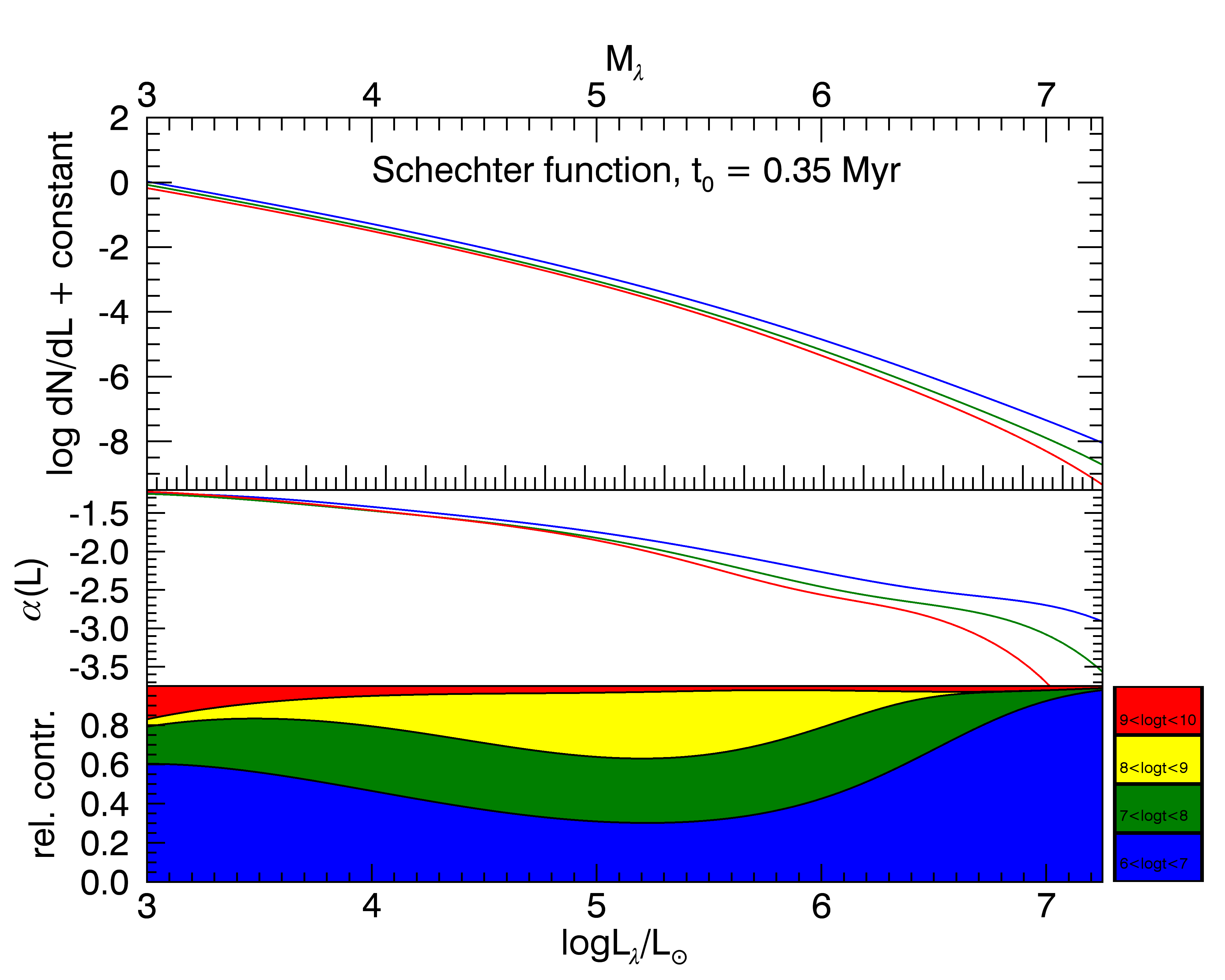}
	\caption{Modelled relative contributions of varying cluster ages made to the luminosity function for a Schechter mass function. The three panels are as follows: the shape of the luminosity function, the value of $\alpha$ for the function at each point  and the contributions of clusters in four different age bands. The blue band is clusters aged $10^6 - 10^7$ years, the green is $10^7 - 10^8$ years, the yellow $10^8 - 10^9$ years and the red $10^9 - 10^{10}$ years. $t_0$ is the average disruption timescale of a 1~$\msun$ cluster.}
	\label{fig:lfmodel}
\end{figure}

\begin{figure}
\includegraphics[width=8.5cm]{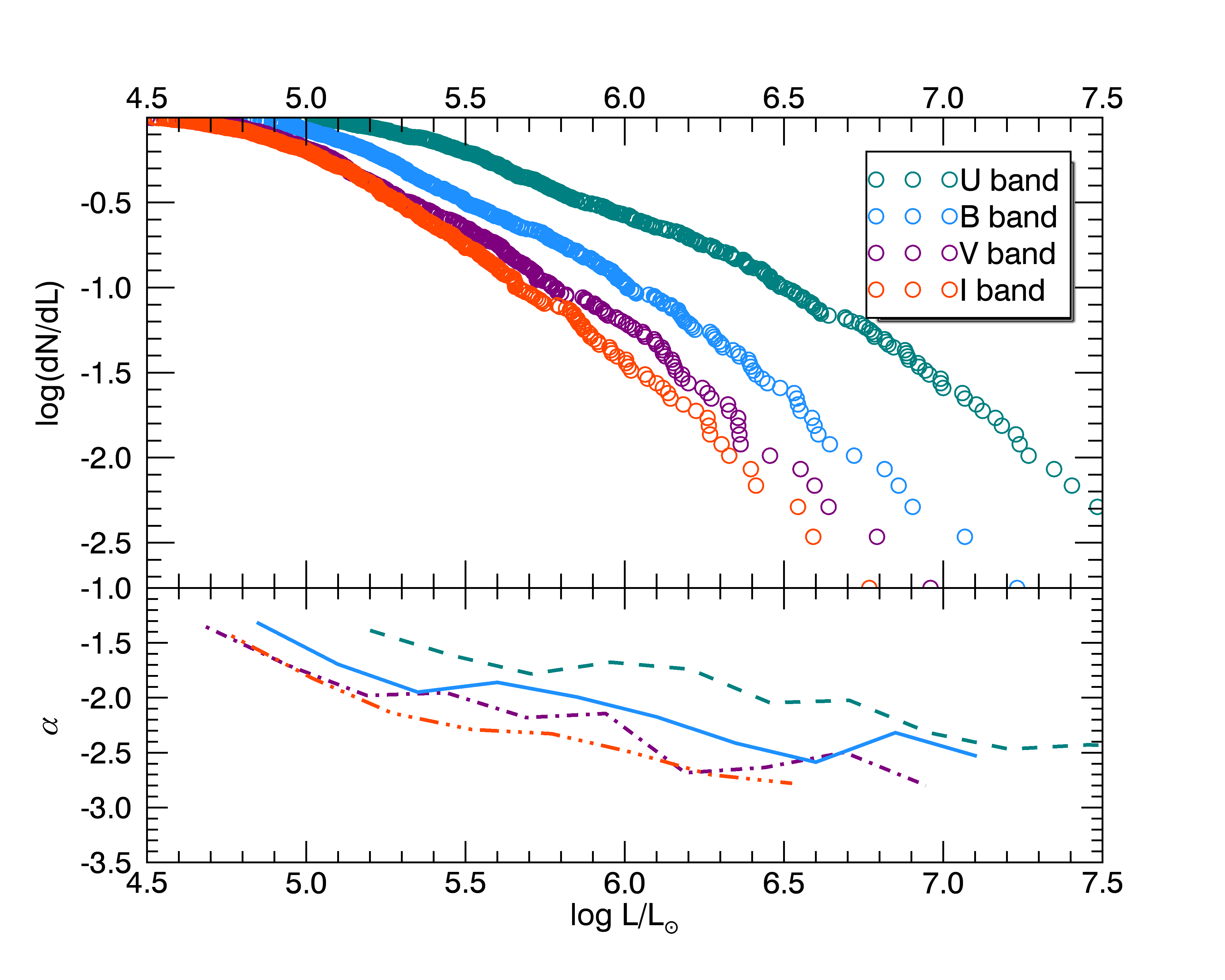}
	\caption{The luminosity functions in the U, B, V and I bands with magnitudes converted to luminosities. The top panel on the plot shows the shape of the functions for all class 1 clusters above 5000 $\msun$ and younger than our incompleteness limit of $\sim 100$ Myr, while the bottom panel shows the fits for the distributions in each band ($\alpha$) along the curves. $\alpha$ is found by fitting the functions above by minimising $\chi^2$, with overlapping bins of 0.5 dex.}
	\label{fig:lfbands} 
\end{figure}

The shape of the luminosity function depends on the cluster formation history, the mass function of the clusters, their mass evolution and extinction. Clusters fade as they age and there are, therefore, clusters with different ages and masses contributing to the number of clusters at a given luminosity. To get a better understanding of how the LF depends on the various implied properties of the cluster population in NGC 1566, we here model the LF based on the underlying mass functions and cluster disruption timescale we derived earlier. 

We assume a constant cluster formation rate of $10^{10}$ yr and a Schechter function for the cluster initial mass function (CIMF), with an index of $\alpha = -2$ at the low-mass end and a truncation mass of $M_* = 2.5\times10^5\,\msun$, as obtained from the maximum likelihood fitting method in \S~\ref{sec:disruption1566}. We use a disruption timescale of $t_0 = 0.35\,$Myr, also as found in \S~\ref{sec:disruption1566}. The mass evolution of individual clusters changes the shape of the cluster mass function. For the mass-dependent cluster disruption model we can use the expression for the `evolved Schechter' mass function given in equation~(26) of \cite{gieles09}. With this we compute the number density of points in age-mass bins, and the contribution of each mass to the LF in the $B$, $V$ and $I$ filters is found from the age-dependent mass-to-light ratios of the single stellar population models of \cite{bruzual03}. The logarithmic slope ($\alpha$) is derived from the LF using the symmetric difference quotient. Similar procedures to model the LF are presented in \cite{fall06, larsen09, gieles10}.

Fig.~\ref{fig:lfmodel} shows the resulting LF model. The top panel shows the LF in the three filters, the middle panel shows the luminosity dependent $\alpha$ and the bottom panel displays the relative contributions of different age bins. Young clusters dominate at the bright-end as the result of the truncation in the CIMF and evolutionary fading (i.e. old $M_*$ clusters are fainter than young $M_*$ clusters). At low luminosities the majority of the clusters are also young, which is due to disruption. 

In Fig.~\ref{fig:lfbands} we show the LF (top panel) and $\alpha$ (bottom panel) for the clusters in NGC 1566. We find $\alpha$ by using the {\sc linfit} procedure in {\sc idl}. We use overlapping bins of 0.5 dex to ensure that no anomalous peaks in the luminosity function dominate the fit for $\alpha$. 

The general behaviour of the LF and $\alpha$ in different filters is similar to that of the model in Fig.~\ref{fig:lfmodel}. The LF is steeper at higher $L$, which in our model is the result of the exponential truncation (at high $L$) and mass dependent disruption (at low $L$). The LF is steeper in redder filters, which in our model is due to the truncation of the mass function and more rapid fading in bluer filters. We note that if the steepening of the LF was due to a luminosity dependent extinction, we would expect to see the opposite: a steeper LF in the bluer filters. The average and median value for $\alpha$ for our data corresponds very well to the average and median for the model values of $\alpha$, as shown in Table~\ref{tab:alphas}.

This modelling exercise further supports the finding of a truncation or steepening in the underlying mass function.

\begin{table}
\centering
{\it Mean values for $\alpha$}
\\
\begin{tabular}{c c c}
Band & Observed & Model \\
\hline
B & -2.08 & -2.02 \\
V & -2.19 & -2.16 \\
I & -2.25 & -2.27 \\
\end{tabular} \\
\vspace{10pt}
{\it Median values for $\alpha$}
\\
\begin{tabular}{c c c}
Band & Observed & Model \\
\hline
B & -1.99 & -1.97 \\
V & -2.14 & -2.11 \\
I & -2.29 & -2.18 \\
\end{tabular}
\caption{Average and median values for $\alpha$ for our observed luminosity function and our model function. The values are very similar across all bands, indicating that our results are comparable.}
\label{tab:alphas}
\end{table}

\section{Galactic parameters and cluster populations}
\label{sec:gamma}

Clusters and associations are the building blocks of galaxies, and therefore are vital to understanding star formation processes. \cite{adamo15} recently showed how tracing the properties of clusters across M 83 provides information on how the galactic environment has influenced the cluster formation process, and consequently gives insight into star formation throughout the galaxy. In this section we discuss several galactic parameters that provide insight into cluster populations and their histories, including $M_V^{brightest}$, $\Gamma$ and $T_L(U)$.

\subsection{Brightest absolute V band cluster}
\label{sec:mvbright}

\begin{figure}
\includegraphics[width=8.5cm]{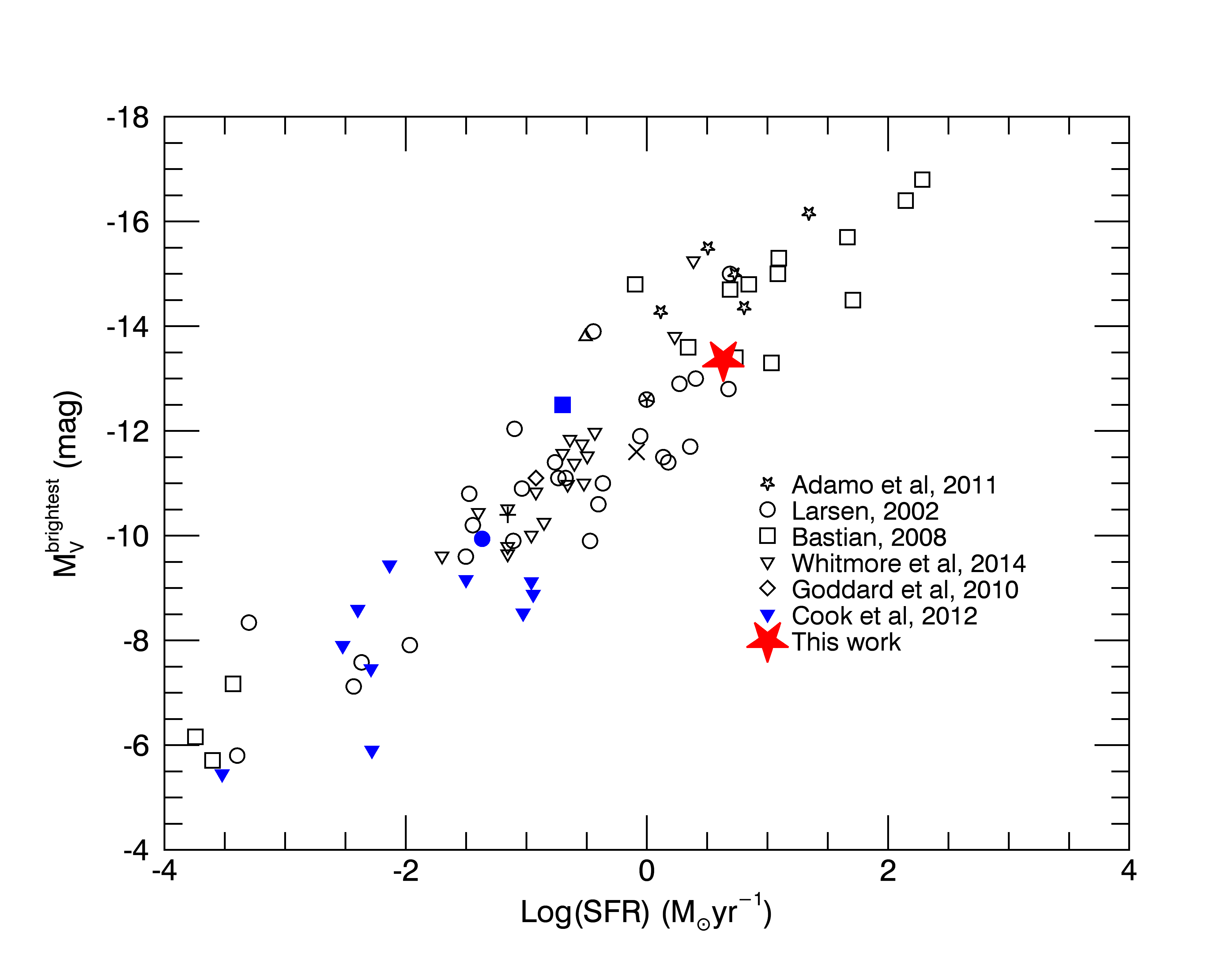}
\caption{A collection of clusters in nearby galaxies' with the brightest V band magnitude in their respective populations plotted against the galaxies' total star formation rates. There is a clear relationship between these two quantities, displaying the stochastic element of the cluster formation process. The value for NGC 1566 is shown as a red star. Major contributors to the number of sources on the plot are listed to the right, please see Table B1 of \protect\cite{adamo15} for the literature containing this information.}
\label{fig:maxmv}
\end{figure}

Fig.~\ref{fig:maxmv} shows the absolute V band magnitude of the brightest cluster in our NGC 1566 catalogue plotted against the log star formation rate of the galaxy. Our SFR was found by reducing the value for the SFR of NGC 1566 provided in \cite{thilker07} to account for the area of the galaxy covered by the HST WFC3 images, which is smaller. \cite{thilker07} use the same distance for NGC 1566  as our study, which makes this calculation trivial. The calculation is based on $H\alpha$ flux extracted from an image covering the entire galaxy from the SINGS survey. The ratio of the flux for the areas covered by \cite{thilker07} and HST allowed us to reduce their estimate of the SFR accordingly, giving us a value of $4.3~\msunyr$. Our data point is displayed as the red star. The other data points for other galaxies were taken from \cite{adamo15} (please see Table B1 of that paper for the full list of objects included). The clear correlation between these two parameters is the result of the stochasticity of the cluster formation process and size of sample effect, where higher SFR galaxies are able to sample the initial mass and luminosity functions to brighter and higher mass clusters \citep{larsen02, bastian08}. NGC 1566 is no exception to this effect, lying comfortably at the end with the higher rate of star formation. 

\subsection{Cluster formation efficiency}
\label{sec:cfe}

As discussed in \S~\ref{sec:props}, the CFE, or $\Gamma$, is the amount of star formation contained within clusters compared to the total star formation in a defined area. Previous studies have found a variation in the CFE among different galaxies when compared to the surface density of star formation ($\Sigma_{SFR}$) (e.g. \citealt{goddard10, ryon14}), with which it correlates. $\Gamma$ is also found to decrease within the same galaxy further from the galactic centre \citep{silvavilla13, adamo15}. As the CFE can provide information on how galactic environment can affect cluster population properties, we calculated the value for NGC 1566 to compare with other galaxies. 

\subsubsection{Calculation method}
\label{sec:gammacalc}

$\Gamma$ is calculated for two different age ranges of clusters: 0-10 Myr and 10-50 Myr. The division at 10 Myr is due to the difficulty in fitting ages and masses to clusters younger than this age, which introduces inaccuracy in the calculation of $\Gamma$ for 0-10 Myr. The mass cut of $5000~\msun$ applied to the rest of this work was again applied here to attempt to account for stochastic IMF sampling by using a mass-limited sample. Additionally, we only included class 1 sources to avoid contamination of a 'true cluster' population. 

$\Gamma$ is found by dividing the cluster formation rate (CFR) by the total star formation rate of the galaxy. So in order to find $\Gamma$ we need to estimate a CFR, which can be defined as the total mass of clusters formed divided by the time over which they form. The first step in calculating a CFR is the integration of a chosen cluster mass function (CMF) over 2 sets of limits. The first set corresponds to finding a theoretical total cluster mass ($M_{c,tot}$), with a lower limit of $100~\msun$ and an upper limit of $1.5\times10^6~\msun$, the observed highest mass cluster. The second set provides an estimate of a theoretical observed cluster mass ($M_{c,obs}$), and uses a lower limit of our mass cut of $5000\msun$ and the same upper limit as the previous integration. Using the ratio of the resulting integrated functions, we can estimate the cluster formation rate in the two selected age ranges and therefore calculate $\Gamma$. 

The CMF we have chosen to use is a simple power law, as observed to fit well to  many cluster populations' mass distributions, including NGC 1566, of the form $dN \propto m^{-\beta} dm$. The validity of using a power law function is discussed in the next section. The value of $\beta$ we chose was 2, the average value found for most galaxies. The ratio of the two integrations provides the factor for calculating a total cluster mass in NGC 1566 based on data, rather than theoretical. For example, if the theoretical ratio of observed to total cluster mass is 0.5, then an estimate of a total cluster mass based on data would be double the summation of all observed clusters in our catalogue. This value is then carried forward to find a CFR by dividing the total mass by 10 for $CFR_{0-10 Myr}$ and 40 for $CFR_{10-50 Myr}$ and $\Gamma$ is then $CFR/SFR\times100$ for each age range.   

\subsubsection{Results and limitations}
\label{sec:gammaresult}      

\begin{figure}
\includegraphics[width=8.5cm]{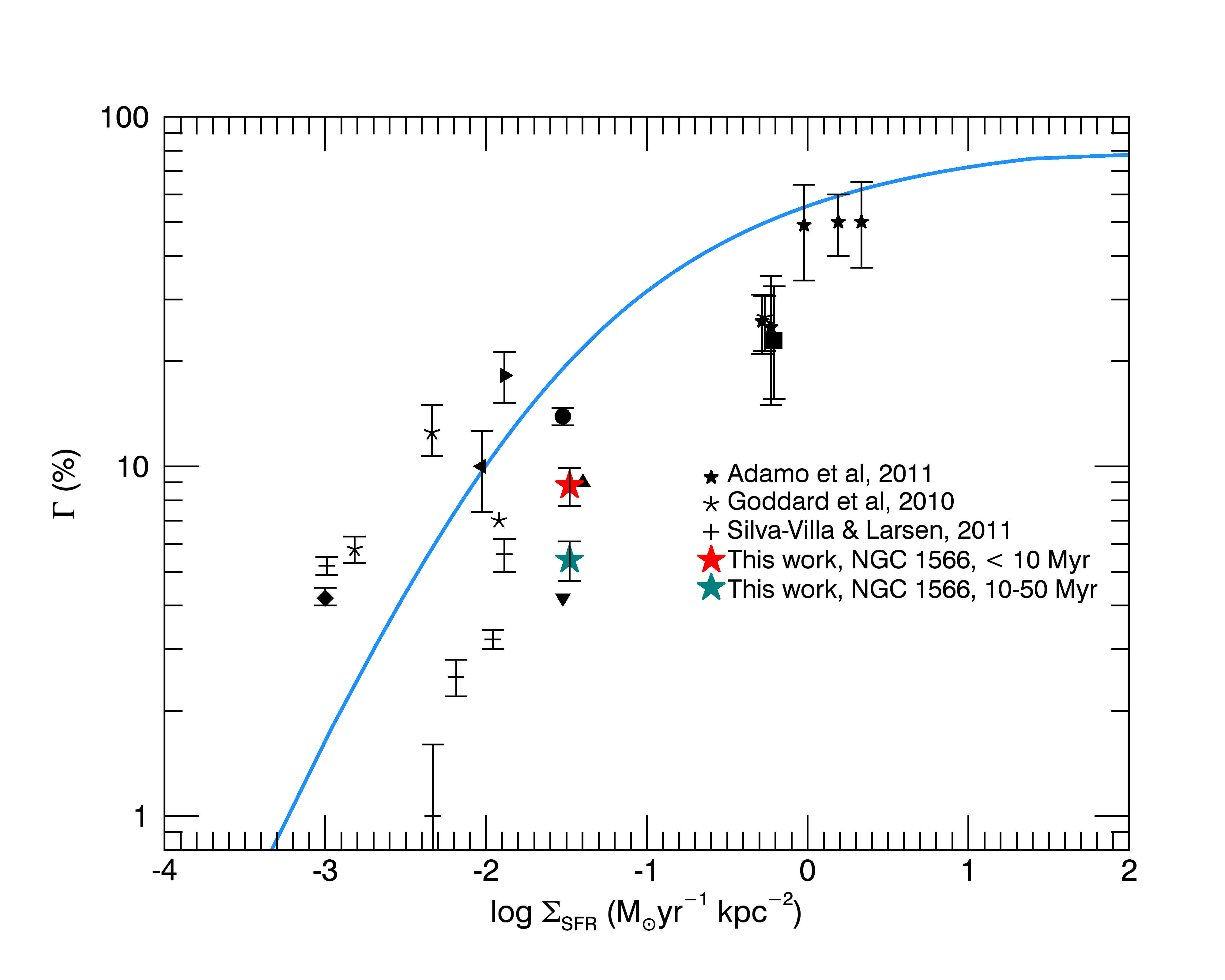}
\caption{Cluster formation efficiency against the surface density of star formation for a variety of objects, including NGC 1566. The correlation indicates that clusters form in denser areas of cold gas. NGC 1566 is shown by the red and teal stars, which display a fairly low CFE and surface density in comparison to the SFR. Major contributors to the data points are listed on the right, the other sources can be found in Table B1 of \protect\cite{adamo15}. The blue line represents the model to the data from \protect\cite{kruijssen12}.}
\label{fig:gamma}
\end{figure}

Fig.~\ref{fig:gamma} shows the trend between $\Gamma$ and log of $\Sigma_{SFR}$ (surface density of star formation) observed for a variety of objects taken from the literature by \cite{adamo15} (again, please see Table B1 in this paper for the full details of the points and their sources), with values calculated for NGC 1566 shown by the red and teal stars. The value of log $\Sigma_{SFR}$ is found by dividing our SFR by the area covered in kpc, and is found to be $\approx 0.033 ~\msun yr^{-1} kpc^{-2}$. For NGC 1566 we find that $\Gamma \approx 8.8 \pm 1.1$ for clusters in the 0-10 Myr age range and $\Gamma \approx 5.4 \pm 0.7$ for the 10-50 Myr range. 

NGC 1566 fits into the correlation with the other points in support of the idea that the amount of star formation occurring in clusters, and therefore the number of clusters is dependent on the surface density of star formation in the galaxy. However, conversely to the idea that a higher SFR can be linked to a higher $\Sigma_{SFR}$ and therefore a higher CFE is that NGC 1566 has a fairly high star formation rate (when compared to similar galaxies) and yet has a lower than expected cluster formation efficiency and $\Sigma_{SFR}$. This indicates that the galaxy is primarily forming stars outside of bound clustered environments possibly because of insufficient gas density. It may be possible, therefore that $\Gamma$ is not strongly related to the star formation rate in the galaxy, and can only be reliably dependent on gas density. Additionally, this could be possible with significant disruption within 100 Myr. 

There are many limitations to our calculation of $\Gamma$, as discussed in detail for NGC 3625 by \cite{goddard10}. In their work they used synthetic cluster populations to examine the accuracy and effectiveness of their calculations. They identified many potential sources of error, which can also be applied to NGC 1566. 

Firstly, by using our total observed cluster mass during the conversion from integrated quantities to estimates based on our data, we assume that we have detected all clusters in the galaxy. Some clusters are inevitably missed during the detection and catalogue refinement phase, especially the youngest clusters that may be obscured by dust. Our catalogue likely misses few of these clusters, however as \cite{whit02} found that when comparing the detection of the youngest clusters in radio bands and optical bands, there is $\sim85\%$ overlap. \cite{me15} found a similar result for M 83, where detections in the H band confirmed few clusters would be missed.

Ages and masses of clusters are obtained by fitting photometric data to SSP models. The parameters of the SSP model, for example the metallicity used, can strongly affect the resulting cluster properties by altering the numbers of clusters at different ages or masses \citep{bastian05}. \cite{goddard10}, however, report a difference of 5-10\% for differing SSP models, so the effect is likely negligible. 

The assumed mass function also plays an important role in the calculation. We have used a power law with an index of $-2$ and an upper limit of two times the observed maximum mass, which fits the majority of the mass function, however, many studies find a Schechter function that incorporates a truncation at the high mass end is a more accurate fit. Our calculation will therefore have overestimated the contribution of high mass clusters to the total amount. An overestimate of the total mass would give a lower value for the CFR and CFE. \cite{goddard10} show that the difference between a power law and Schechter function makes little difference in the calculation of the total mass for a truncation value of $10^6~\msun$ or higher. The estimate of the truncation value for NGC 1566, however, is $\approx 10^{5.4}~\msun$, which is shown to have a larger difference. The difference in integrated mass between a power law and a Schechter function with turnover mass equal to that of NGC 1566 is $\approx 0.75$. This means that our total mass could be $\approx 1.3$ times lower than we calculated, giving a value for $\Gamma$ that is also 1.3 times lower, or $\approx 6.8$ for 0-10 Myr and $\approx 4.2$ for 10-50 Myr.  

\cite{chandar15} recently presented a new statistic mean to test the CFE; a relation involving the cluster mass function normalised by the star formation rate of the galaxy (CMF/SFR).  They report only a weak correlation between the CMF/SFR and the SFR of the host galaxy, and no correlation with $\Sigma_{SFR}$ for the young ($<$10 Myr) cluster population in their sample.  When using an older population (100-300 Myr), they did find similar trends as expected based on previous works using $\Gamma$ (e.g. \citealt{adamo15}). Their CMF/SFR is not subject to all of the same uncertainties as $\Gamma$, though ages, masses and extinctions do still need to be modelled. \cite{kruijssen16} showed that the discrepancy, at least in part is due to a lack of distinction of bound and unbound aggregates at young ages in \cite{chandar15} as well as the need to account for cluster disruption at later ages. It is worth noting, however, that some of the galaxies presented in \cite{adamo15} also do not make the distinction between bound and unbound aggregates, though several make age cuts to remove young clusters that likely cause contamination of unbound sources. This lack of uniformity is addressed by the LEGUS survey \citep{calzetti15}.

\subsection{$T_L(U)$}
\label{sec:tlu} 

Another quantity that is related to $\Gamma$ and gives an indication of the effect that environment plays on cluster populations is the percentage of U band light from a galaxy that is emitted by clusters, or $T_L(U)$. U band light primarily traces the young clusters in the population, as they are usually brighter in bluer bands, while older clusters have faded and emit more strongly in redder bands, so therefore should be linked to star formation. Unlike $\Gamma$ however, $T_L(U)$ is a purely observational quantity, therefore free of the biases and errors introduced by selecting an approximate mass function and using quantities derived from SSP models. $T_L(U)$ is also not strongly affected by extinction.

We took the data for other galaxies from \cite{larsen00} and \cite{adamo11} and calculated $T_L(U)$ for NGC 1566 to investigate where it lay in relation to other galaxies on the plot. $T_L(U)$ was calculated by summing the U band luminosities of all clusters in the catalogue, dividing by the total measured U band luminosity for the entire galaxy obtained by aperture photometry and multiplying by 100. We applied the usual mass cut of $5000~\msun$ and split the catalogue into class 1 and class 2 sources. We find that $T_L(U) \approx 10.1\%$ for class 1 sources and $T_L(U) \approx 12.7\%$ for class 1 and 2 sources.

\begin{figure}
\includegraphics[width=8.5cm]{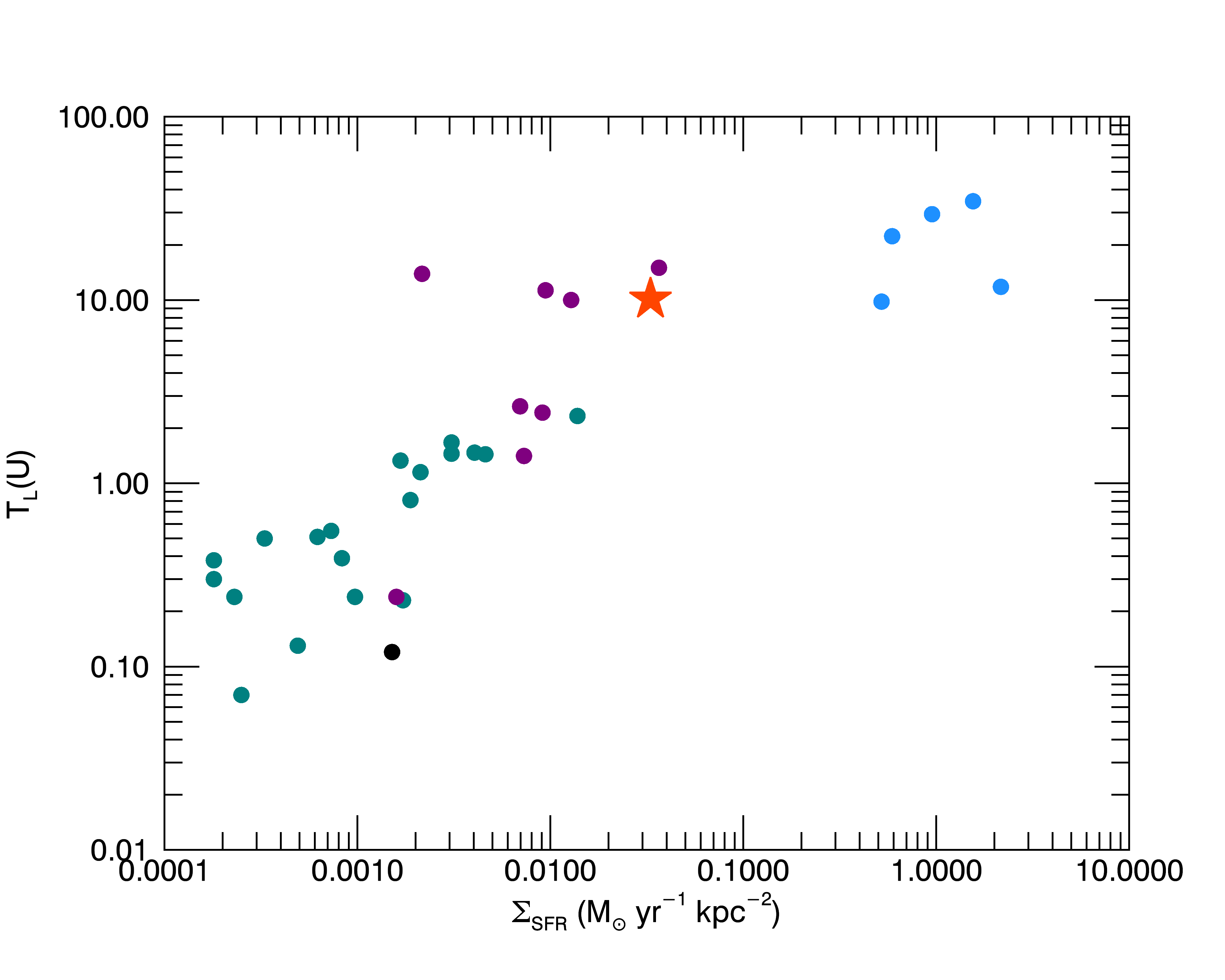}
\caption{$T_L(U)$ for galaxies taken from \protect\cite{larsen00} and \protect\cite{adamo11} and now including the data for NGC 1566. The teal points indicate galaxies that were taken from \protect\cite{larsen1999}, the purple points are starburst and merger galaxies introduced in \protect\cite{larsen00}, the black points are other galaxies from \protect\cite{larsen00} and the blue are BCG galaxies from \protect\cite{adamo11}. The red star is our data for NGC 1566. Both axes are log units.}
\label{fig:tul} 
\end{figure}

Fig.~\ref{fig:tul} shows the relationship between $\Sigma_{SFR}$ and $T_L(U)$. The purple, teal and black points are those taken from \cite{larsen00}, where purple points are starburst galaxies and mergers, while the black and teal points are other galaxies. The blue points are BCGs taken from \cite{adamo11} and the red star is for NGC 1566. The plot demonstrates that the galaxy fits well onto the current relationship and is also in the section of the plot populated by starburst galaxies and merging systems. This would be expected of a galaxy with a high star formation rate, though $T_L(U)$ correlates less strongly with SFR than $\Sigma_{SFR}$ \citep{larsen00, larsen02}.

\begin{figure}
\includegraphics[width=8.5cm]{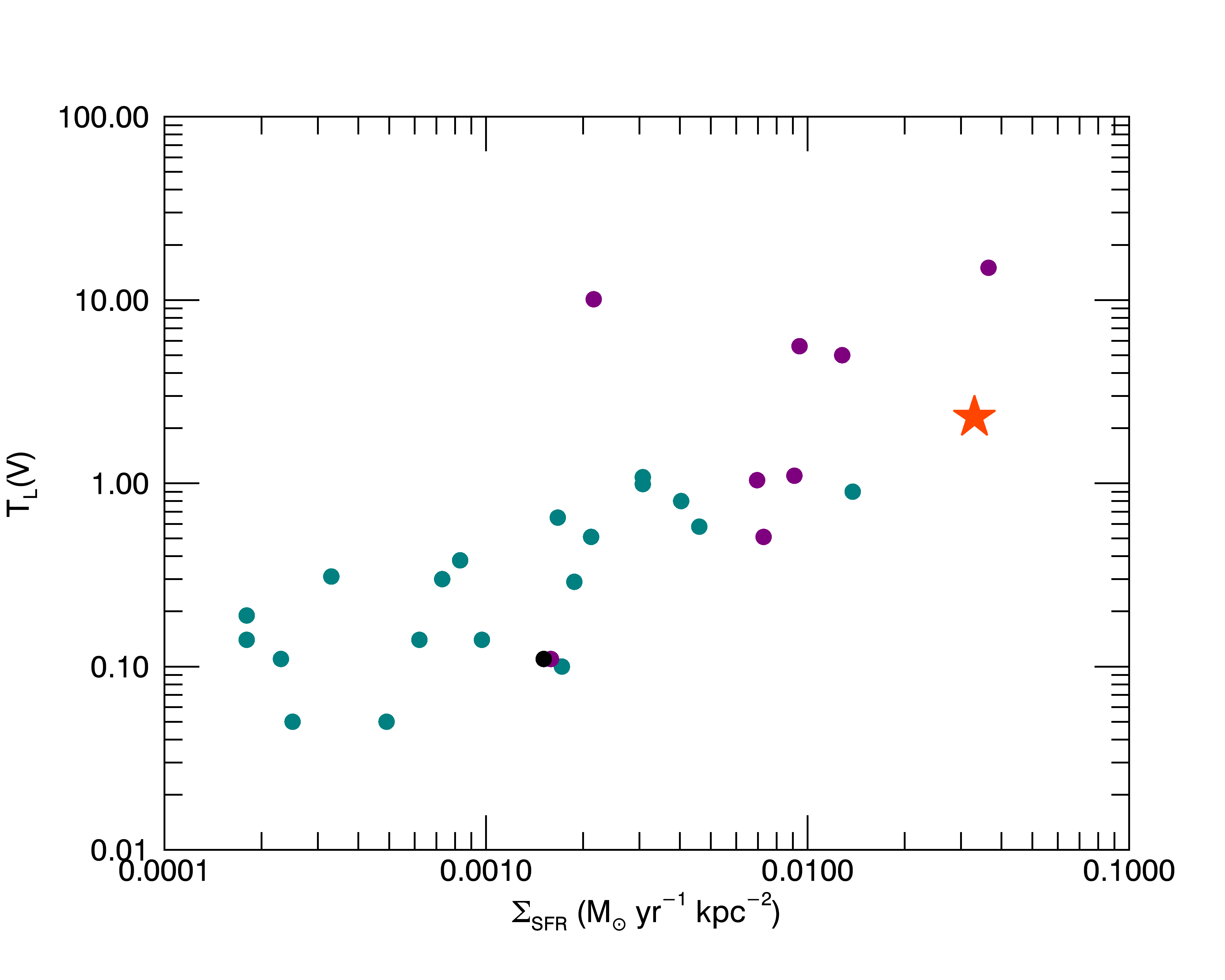}
\caption{$T_L(V)$ for galaxies taken from \protect\cite{larsen00} and now including the data for NGC 1566. The colours indicate the same separation of galaxies as described in the previous figure.}
\label{fig:tvl} 
\end{figure}

In addition to $T_L(U)$, \cite{larsen00} provided values for $T_L(V)$ for all of the galaxies. Fig.~\ref{fig:tvl} shows the relationship of $T_L(V)$ with $\Sigma_{SFR}$. Little difference is seen between the plots for the two different bands. NGC 1566 however, shows a fairly large difference between the U and V bands. If the star formation history has been increasing, this could be due to the galaxy having many young clusters, so they contribute more strongly to $T_L(U)$ than $T_L(V)$. We investigated the effect on the percentage of the total luminosity emitted from clusters in the U, B, V and I bands. 

\begin{table}
\centering
\begin{tabular}{c c c c}
Band & $M_{gal}$ & $T_L^1$ & $T_L^{1+2}$ \\
\hline
U & 11.59 & 10.1 & 12.7 \\
B & 11.81 & 4.8 & 5.9 \\
V & 10.94 & 2.3 & 2.8 \\
I & 9.76 & 1.3 & 1.5 \\
\end{tabular}
\caption{Percentage of total galaxy luminosity emitted from clusters in the U, B, V and I bands. $M_{gal}$ is the magnitude of the whole galaxy in each band, $T_L^1$ is the percentage given for only class 1 sources and $T_L^{1+2}$ is for class 1 and 2 sources.}
\label{tab:tuldata}
\end{table}

Table~\ref{tab:tuldata} shows the data for the different bands, giving the total magnitude of the galaxy and the percentages for only class 1 sources and class 1 and 2 sources. There appears to be a general trend, as we could expect, in the values of $T_L$. Less luminosity in redder bands is omitted by clusters, as older clusters that emit primarily at redder colours are fainter and will contribute less than other sources. This indicates that by calculating $T_L(I)$ we are comparing the light from clusters to the large field stellar population throughout the galaxy, consisting of older stars, which will contribute much more strongly at longer wavelengths. Younger clusters are usually the brightest, which emit more heavily in bluer bands.  

\section{Conclusions}
\label{sec:discussion}

\subsection{Radial variation in cluster properties}
\label{sec:clusterprops}

\begin{table}
\begin{tabular}{c c c c}
\centering
 & Radial bin 1 & Radial bin 2 & Radial bin 3 \\
 \hline
$N_1$ & 274 & 274 & 278 \\
$N_2$ & 70 & 85 & 56 \\
$\alpha_V$ & 2.09 & 2.15 & 2.04 \\
$\beta$ & 1.79 & 1.96 & 2.00 \\
$M_c$ (\msun) & $2\times10^5$ & $3\times10^5$ & $2\times10^5$ \\
$t_4$ (Myr) & 80 & 80 & 200 \\
\end{tabular}
\caption{Summary of data for the three radial bins used throughout this study. These are values as calculated using a mass cut of 5000 \msun\ with full detections in the UV band. The bins were chosen to accommodate approximately equal numbers of clusters in each bin for our plots. However, this applies only to class 1 sources, as these were used for the analysis. $N_1$ and $N_2$ refer to the number of class 1 and class 2 sources respectively. $\alpha_V$ is the fit to the V band luminosity function, while $\beta$ is the fit to the mass function. $M_c$ is the truncation mass and $t_4$ is the average disruption timescale of a $10^4~\msun$ cluster.}
\label{tab:summary}
\end{table}

NGC 1566 displays negligible variations in several cluster properties at different galactocentric distances within the $\sim5.5$~kpc covered by the WFC3 observations. A summary of the values calculated for various properties for each bin is shown in Table~\ref{tab:summary}. The conclusions we have reached from this study are as follows:

\begin{itemize}
\item{There is a small variation in colour space with respect to galactocentric distance. The most concentrated areas of colour space are slightly redder in the outer radial bins. This is likely due to small variations in the age of the clusters with distance, as age is the primary factor affecting colour distribution (extinction is expected to go in the opposite direction). Similar variations are also seen in other galaxies, such as NGC 4041 and M 83.}

\item{The shape of the luminosity functions for all radial bins are as expected, with the power law section of the best fit with an index value of $\alpha\approx-2$, as found in numerous other studies and galaxies. We found negligible differences in the indices fitted for different radial bins, with the largest differences in the UV and U bands, though all are within errors on the fit. In agreement with the luminosity function, as they are potentially related, the mass distributions also show only small variations. We find a steepening of $\alpha$ for redder bands, which is predicted if the cluster mass function is truncated at the end mass end.  This is  due to the more rapid fading of clusters in bluer bands, as clusters with the same mass, but different ages, are more spread across a luminosity range in bluer bands, giving rise to a shallower function.}  

\item{The age distribution for NGC 1566 is also of the shape we would expect. Little difference in colour distribution for clusters in the three radial bins indicated a fairly uniform range of cluster ages throughout the galaxy. The age distribution shows that this is likely the case, as the three bins display only small variations between them. 
There is some difference between the shape of the distribution for NGC 1566 and other galaxies such as M~31 and M~83.  The inner regions of M~83 have a much steeper age distribution between $10-100$~Myr, suggesting that cluster disruption is much more efficient there than in NGC~1566 or M~31. }

\item{The mass distributions show a slight steepening with increasing radial distance, but within error estimates.  Additionally, the data suggest a truncation in the mass function, though this could be due to low numbers of clusters at high masses. This finding is supported by our comparison of the observed luminosity function with models, which show that a Schechter function is a good fit. The index of the slopes are all $\sim 2$, as observed for other galaxies.}

\end{itemize}

\subsection{Galactic properties}
\label{sec:galconc}

We have calculated $T_L(U)$ and $\Gamma$ for NGC 1566 and find both values to lie within the correlations with $\Sigma_{SFR}$. An interesting result of the study was that we found $\Gamma$ to be slightly lower than expected for a galaxy with a fairly high SFR. This is when compared to galaxies similar to NGC 1566. While the galaxy still fits into the current scaling relations for these properties, $\Gamma$ could indicate that the galaxy is less efficient at forming stars in clusters than we may expect.

\section*{Acknowledgements}

Based on observations made with the NASA/ESA Hubble Space Telescope, and obtained from the Hubble Legacy Archive, which is a collaboration between the Space Telescope Science Institute (STScI/NASA), the Space Telescope European Coordinating Facility (ST-ECF/ESA) and the Canadian Astronomy Data Centre (CADC/NRC/CSA).

\bibliographystyle{mn2e}

\bibliography{refs2}

\label{lastpage}
\end{document}